\documentclass[twocolumn,aps,groupedaddress]{revtex4}

\usepackage{graphicx}
\usepackage{dcolumn}
\usepackage{bm}
\usepackage{amsmath,amssymb,amsfonts}
\usepackage{color}
\usepackage{ulem}
\usepackage[caption=false]{subfig}

\begin{document}

\title{Magic of high-order van Hove singularity}
\author{Noah F. Q. Yuan}
\author{Hiroki Isobe}
\author{Liang Fu}
\affiliation{Department of Physics, Massachusetts Institute of Technology, Cambridge, Massachusetts 02139, USA}

\begin{abstract}
We introduce a new type of van Hove singularity in two dimensions, where a saddle point in momentum space is changed from second-order to high-order. Correspondingly, the density of states near such ``high-order van Hove singularity'' is significantly enhanced from logarithmic to power-law divergence, which promises stronger electron correlation effects. High-order van Hove singularity can be generally achieved by tuning the band structure with a single parameter in moir\'e superlattices, such as {twisted bilayer graphene by tuning twist angle or applying pressure, and trilayer graphene by applying vertical electric field}.
\end{abstract}

\maketitle


The recent discovery of unconventional insulating states and  superconductivity in twisted bilayer graphene (TBG) \cite{exp1,exp2} has attracted enormous interest. At ambient pressure, these intriguing phenomena of correlated electrons only occur near a specific  twist angle $\theta \approx 1.1^\circ$, widely referred to as the magic angle. The existence of such a magic angle was first predicted from  band structure calculations. Based on a continuum model \cite{Neto, Neto2}, Bistritzer and MacDonald  showed \cite{MacDonald} that the lowest moir\'e bands in TBG become exceptionally flat at this magic angle, and are hence expected to exhibit strong electron correlation. This pioneering work inspired a large body of experimental works in recent years, which culminated in the discovery reported in Ref. \cite{exp1,exp2}. The origin of magic-angle phenomena, i.e., the emergence of superconductivity and correlated insulator, is now under intensive study \cite{Xu,Volovik,Yuan,Po1,Kivelson,Phillips,Lee,Oskar,Thomson,Biao,Venderbos,Fidrysiak,
Ying,Guinea,Ochi,XCWu,FCWu,FCWu1,Ray,Zhang}.

While the reduction of bandwidth is undoubtedly important, it is evident that the phenomenology of magic-angle TBG cannot be ascribed to a completely flat band. Quantum oscillation at low magnetic fields reveals doping-dependent Fermi surfaces {\it within} the narrow moir\'e band.
{Detailed}  scanning tunneling spectroscopy (STS) studies \cite{Abhay, Stevan, Yazdani, LeRoy} {and compressibility measurements \cite{Ashoori}} show that the moir\'e bandwidth at a magic angle is a few tens of meV, larger than what previous calculations found \cite{MacDonald,NamKoshino, KoshinoFu, Shiang, BS1,BS2,BS3}.
Moreover, there is no {direct} evidence from existing STS measurements {at various twist angles} that the moir\'e band is {exceptionally flat right at} the magic angle (see also Ref. \cite{LeRoy}). These experimental results motivated us to consider additional feature of electronic structure which may create favorable condition for correlated electron phenomena in magic-angle TBG.

\begin{figure}
\includegraphics[width=0.5\textwidth]{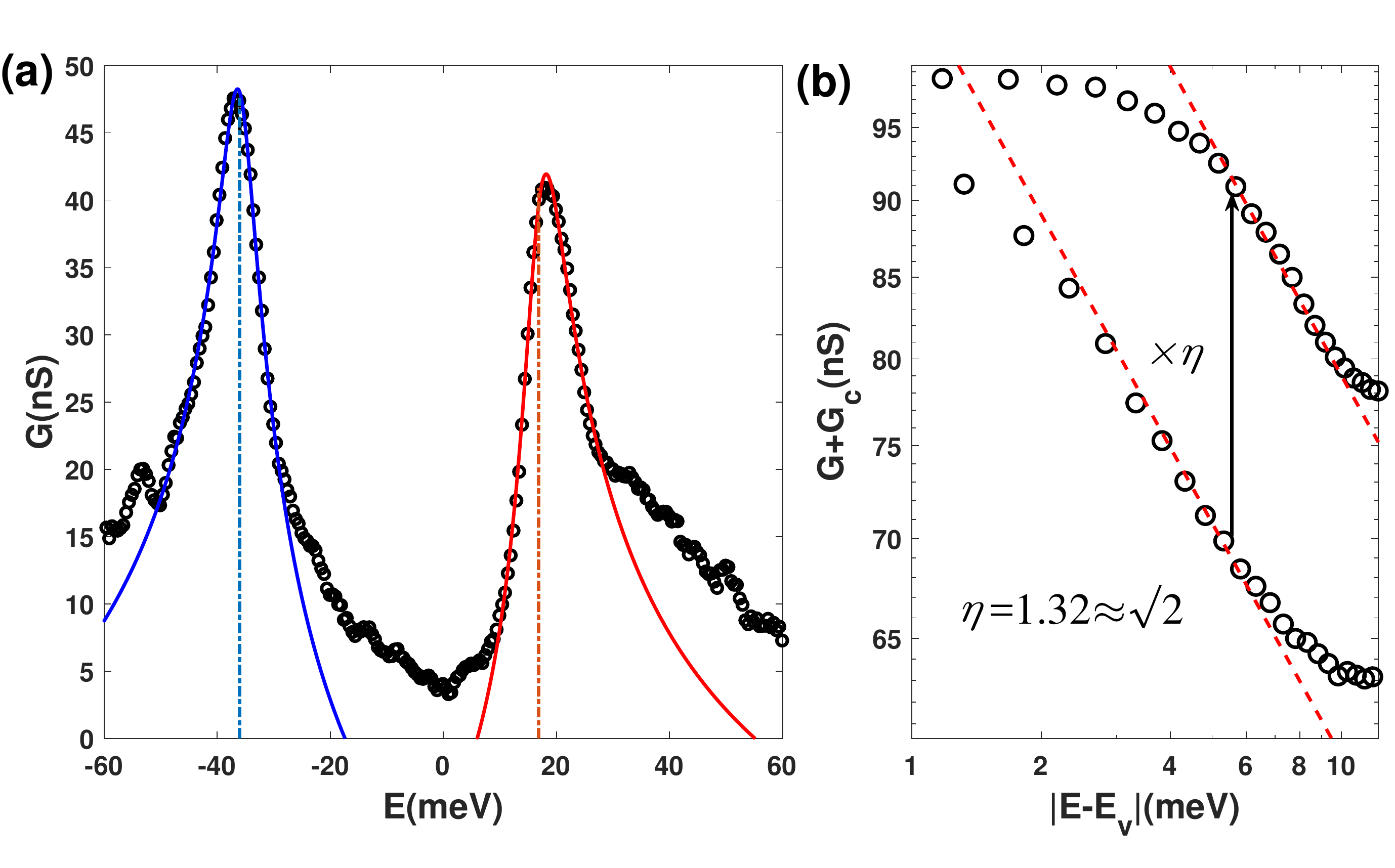}
\centering
\caption{{Theoretical fit to the tunneling conductance measurement \cite{Abhay}.} (a) {Open circles are tunneling conductance $ G $ of twisted bilayer graphene at twist angle $1.10^\circ$, and solid lines are fitting according to Eq. (\ref{eq_DOS}) with details given in the Supplemental Material. Dashed lines denote singularity energies, indicating the asymmetry of conductance peaks. (b) The peak {at $E_{\rm v}=16.72$ meV} plotted in logarithmic scales. Electron and hole sides of the peak fall into two parallel lines with the same slope $ -1/4 $ and asymmetry ratio $ \eta=1.32\approx\sqrt{2} $. The only parameter in (b) is the background offset $ G_c=57.6 $ nS.}}\label{fig_0}
\end{figure}

In this work, we propose a new perspective that relates magic angle to a new type of van Hove singularity (VHS) in the single-particle energy spectrum of TBG. The importance of VHS has already been recognized in our theory of correlated TBG \cite{Isobe, Kozii} and related studies \cite{Nandkishore, VH1,VH2,VH3,VH4,VH5,VH6} using the weak coupling approach. Generally speaking, VHS in two-dimensional systems are associated with saddle points of energy dispersion in $\bm k$ space.
When a VHS is close to Fermi energy, the increased density of states (DOS) amplifies electron correlation, resulting in various ordering instabilities such as density wave and superconductivity at low temperature \cite{Isobe}.
{Indeed, the two recent STS measurements on gate-tunable TBG around magic angle \cite{Abhay, Stevan} find that} when the VHS shifts to the Fermi level under gating (the corresponding density is within 10\% of half filling), the VHS peak in tunneling density of states splits into two new peaks (see also Ref. \cite{Lin}).
These {findings} clearly demonstrate the prominent role of VHS in magic-angle TBG.

On the other hand, it is {also} known from general consideration and previous STS measurements \cite{stm1,stm2,stm3} that VHS are present at all twist angles. It is therefore unclear whether VHS at magic angle is anything special.

We propose the following scenario. As the twist angle decreases below a critical value $\theta_c$, the van Hove saddle point---which marks the change of topology in Fermi surface (Lifshitz transition)---undergoes a topological transition whereby a single saddle point 
splits into two new ones. Right at $\theta_c$, the saddle point changes from second-order to higher-order; the DOS at VHS is significantly enhanced from logarithmic to power-law divergence, which promises stronger electron correlation. We propose that proximity to such ``high-order van Hove singularity'', which requires tuning to the critical {twist} angle or pressure, is {an} important factor responsible for correlated electron phenomena in TBG near half filling.

We demonstrate by topological argument that high-order VHS can be {\it generally} achieved by tuning the band structure with just a single {control} parameter.
For TBG, besides twist angle, pressure can also induce superconductivity at a twist angle larger than $1.1^\circ$ \cite{Dean}. By tracking the evolution of the moir\'e band with twist angle and pressure, we locate the high-order VHS in experimentally relevant parameter regime and predict its key feature, a distinctive {\it asymmetric} peak in DOS. This feature compares well with the experimentally observed VHS peak in magic-angle TBG, {as shown in Fig. \ref{fig_0}}. We also {discuss} the splitting of VHS peak by nematic or density wave order.

\begin{figure*}
\includegraphics[width=\textwidth]{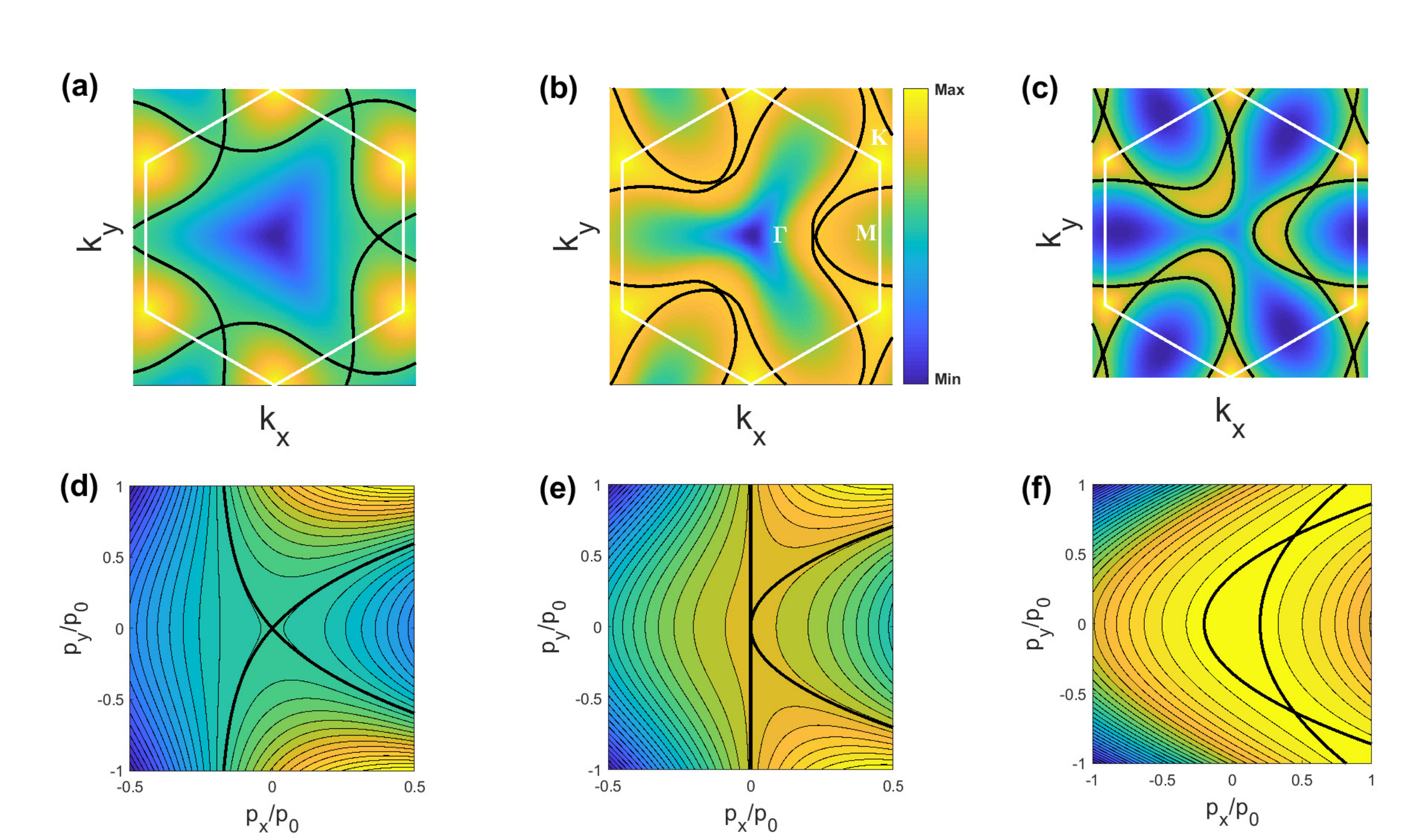}
\centering
\caption{{Energy contours of continuum model in (a-c) and polynomial dispersion (\ref{expand}) in (d-f), and thick lines are energy contours at VHS energy $ E_{\rm v} $. In continuum model $ g'=1.2g $ with (a) $g=1$ (b) $g=1.9$ and (c) $ g=2 $. In polynomial dispersion $ p_{0}=\alpha/\tilde{\gamma} $ with parameters (d) $ \beta =0.2\alpha,\kappa=0 $, (e) $ \beta =0,\kappa=0 $ and (f) $ \beta =-0.2\alpha,\kappa=-0.2\gamma^2/\alpha $.}}\label{fig_1}
\end{figure*}

{\it Ordinary and high-order VHS. }
In two-dimensional (2D) electron systems with energy dispersion $E(\bm k) $, an ordinary VHS with logarithmically diverging DOS occurs at {a saddle point ${\bm k}_s$, determined by}
\begin{eqnarray}
{%
\nabla_{\bm k} E(\bm{k}_s)  =\bm 0 \text{ and } \det D<0,
}
\end{eqnarray}
where $ D_{ij} \equiv \frac{1}{2}\partial_i\partial_j E $ is the $2\times 2$ Hessian matrix of $E$ at ${\bm k}_s$. Since $D$ is symmetric by definition, we can rotate the axes to diagonalize $D$ and Taylor expand the energy dispersion near ${\bm k}_s$ as
$E-E_{\rm v}=-\alpha p_{x}^2+\beta p_{y}^2$,
where $ E_{\rm v} $ is the VHS energy, {the momentum $\bm p={\bm k} -{\bm k}_s$ is measured from the saddle point, and the coefficients $-\alpha$, $\beta$} are the two eigenvalues of $D$ with $ -\alpha\beta =\det D<0 $. This dispersion describes {\it two} pieces of Fermi contours that intersect at the saddle point ${\bm k}_s$.
It is known from a topological consideration that due to the periodicity of $E(\bm k) $ on the Brillouin zone (a torus), van Hove saddle points appear quite generally in energy bands of two-dimensional materials \cite{vanHove}.

We now introduce the notion of high-order VHS by the following conditions {at a saddle point}
\begin{eqnarray}
{%
\nabla_{\bm k} E(\bm{k}_s)  =\bm 0 \text{ and } \det D=0,
}
\end{eqnarray}
which imply $ \alpha\beta =0 $.
There exist two types of high-order VHS. The first type corresponds to $ \alpha =\beta =0 $, i.e., $ D=\bm 0 $. The Taylor expansion of $E(\bm k) $ around such saddle point then starts from at least the third order, and describes an intersection of three or more Fermi surfaces at a common $\bm k$ point \cite{FuSachdev}. Such high-order VHS (type-I) is also referred to as  \textit{multicritical} VHS in the recent literature \cite{multicritical1,multicritical2, ruthenate}.
 
Here we focus instead on a new type of high-order VHS relevant for TBG, where the Hessian matrix $D$ has a single zero eigenvalue ($\beta=0$), while the other eigenvalue is nonzero  ($\alpha\neq 0$).
Then in the Taylor expansion of $E(\bm k) $, besides the single second-order term $\alpha$ we must also include higher order terms in order to capture the Fermi contours nearby.
Importantly, unlike the previous case, type-II high-order VHS still describes the touching of two Fermi surfaces, but they touch tangentially (generally speaking) rather than intersect at a finite angle as in the case of ordinary VHS.

{\it VHS in TBG.}
We now turn to VHS in TBG, whose low-energy moir\'e bands arise from inter-layer coupling of Dirac states on the two graphene layers.
Since single-particle scattering between $\bm K$ and $\bm K'$ points requires large momentum transfer, it is suppressed at small twist angles due to the long wavelength of moir\'e potential. In the absence of valley hybridization, the moir\'e bands from $\bm K$ and $\bm K'$ valleys are decoupled \cite{Neto,Neto2,MacDonald, Pablo}, with energy dispersions denoted by $E_+(\bm k)$ and $E_-(\bm k)$ respectively.   Time-reversal symmetry implies $E_-(\bm k) = E_+(-\bm k)$, so it suffices to consider $E_+(\bm k)$ only in the following.

The number and location of VHS points in TBG depend on the twist angle. For example, the band structure calculation for $\theta=2^\circ$ \cite{smet}   reveals  
the existence of three symmetry-related van Hove saddle points on $\Gamma M$ lines in the mini-Brillouin zone (MBZ), shown in Fig. \ref{fig_1}a. Across the VHS energy, the Fermi contour changes from two disjoint Dirac pockets around MBZ corners to a single pocket enclosing the MBZ center. This leads to the conversion between electron and hole charge carriers, as evidenced by the sign change of Hall coefficient \cite{smet}.
On the other hand, at $\theta=1.05^\circ$,  on each  $\Gamma M$ line there is a local energy maximum---instead of saddle point---in the moir\'e valence band. As doping increases, an additional hole pocket (not present at $\theta=2^\circ$) emerges out of each energy maximum and eventually intersects two Dirac pockets at two new saddle points on opposite sides of $\Gamma M$ \cite{KoshinoFu}. In such band structure there are a total of 6 VHS points, shown in Fig. \ref{fig_1}c.

By continuity, we deduce from this change in the number of VHS points that a topological transition of saddle points must occur, at a hitherto unknown critical twist angle, in such a way that a saddle point on $\Gamma M$ (denoted by $\Lambda_0$)  splits into a pair of new ones off $\Gamma M$ (denoted by $\Lambda_\pm$).
Importantly, the behavior of these VHS points in the vicinity of this transition is solely governed by the {\it local} energy dispersion near $\Lambda_0$.

We now expand $E(\bm k)$ near $\Lambda_0$ to higher orders:
\begin{eqnarray}\label{expand}
E-E_{\rm v}=-\alpha p_{x}^2+\beta p_{y}^2+\gamma p_{x}p_{y}^2 +\kappa p_{y}^4 +\ldots,
\end{eqnarray}
where $p_x$ ($p_y$) is parallel (perpendicular) to $\Gamma M$ line.
In the expansion (\ref{expand}), first-order terms vanish because $\Lambda_0$ has by definition zero Fermi velocity. Moreover, only even power terms of $p_y$ are allowed because of the two-fold rotation symmetry of TBG, which acts within each valley and maps $(p_x, p_y)$ to $(p_x, -p_y)$. 
The third-order $\gamma$ term and fourth-order $ \kappa $ term are essential to describe the splitting of VHS across the transition, {which we shall show below along with the relation to scaling properties.}

The behavior of VHS and Fermi contour of the energy dispersion (\ref{expand}) depends crucially on the sign of $\alpha\beta$. Without loss of generality we assume {$\alpha>0$} in the following. For $\beta>0$, ${\bm p}=\bm 0$ (i.e., $\Lambda_0$) is an ordinary van Hove saddle point with logarithmic divergent DOS. The Fermi contour at the VHS energy consists of two curves that approaches straight lines  $p_y/p_x = \pm \sqrt{\alpha/\beta}$ as $\bm p \rightarrow \bm 0$, intersecting at a finite angle $\varphi = 2 \arctan(\sqrt{\beta/\alpha})$ as shown in Fig. \ref{fig_1}d. This behavior corresponds to the VHS in the calculated band structure of TBG at $\theta = 2^\circ$.
When $ \beta\to 0^{+} $, $ \varphi\to 0 $ so that the two Fermi contours at the VHS energy tend to touch tangentially.
On the other hand, for $\beta<0$, $\bm p=\bm 0$ becomes a local energy maximum, while two new saddle points appear at momenta
\begin{eqnarray}
{\Lambda}_\pm = (- \beta\gamma/\tilde{\gamma}^2 , \pm \sqrt{-2\alpha \beta}/\tilde{\gamma})
\end{eqnarray}
whose energy is shifted from $E_{\rm v}$ by $\delta = -\alpha\beta^2/\tilde{\gamma}^2 <0$,  where $ \tilde{\gamma}\equiv\sqrt{\gamma^2 +4\alpha\kappa} $.
Throughout this manuscript we consider the regime $ \gamma^2 +4\alpha\kappa> 0 $ so that $ \tilde{\gamma} $ is always real and positive, and two split saddle points $ \Lambda_{\pm} $ are well-defined.
$\Lambda_+$ and $\Lambda_-$ are a pair of ordinary VHS points related by two-fold rotation, whose Fermi contours are two parabolas $ 2\alpha (p_x\mp \beta/\tilde{\gamma})=(\gamma\pm\tilde{\gamma}) p_y^2$, see Fig. \ref{fig_1}f.  This behavior corresponds to the VHS in the calculated band structure at $\theta = 1.05^\circ$. In the limit $\beta \rightarrow 0^-$, $\Lambda_\pm$ approaches $\Lambda_0$.

Our unified description of two regimes of ordinary VHS ($\Lambda_0$ and $\Lambda_\pm$) using a single local energy dispersion (\ref{expand}) implies that the transition between them corresponds to the sign change of $\beta$. In this process, energy contours at VHS changes from intersecting at one point $\Lambda_0$ (Fig. \ref{fig_1}d and e) to two points $\Lambda_\pm$ (Fig. \ref{fig_1}f). Right at $\beta=0$, $\bm p = \bm 0$ becomes a high-order VHS point. The Fermi contour in its vicinity consists of two parabolas $2\alpha p_x=(\gamma\pm\tilde{\gamma}) p_y^2$, touching each other tangentially at $\bm p=\bm 0$.
{In the special case with $\kappa=0$, one of the parabolas becomes a straight line, as shown in Fig. \ref{fig_1}e.}

The DOS can be used as an indicator of this topological transition of VHS.
For the energy dispersion (\ref{expand}), we find the DOS analytically
{
\begin{widetext}
\begin{eqnarray}\label{eq_DOS}
\rho(E) = \frac{-1}{\pi}{\rm Im}\int\frac{d^2\bm p}{(2\pi)^2} \frac{1}{E+i0^{+}-E(\bm p)} =\frac{\text{sgn}(\alpha \beta)}{\sqrt{2}\alpha\pi^{2}}{\rm Re}\left[\frac{1}{\sqrt{z_{-}}}{\rm K}\left(1-\frac{z_{+}}{z_{-}}\right)-\frac{2i}{\sqrt{z_{+}}}{\rm K}\left(\frac{z_{-}}{z_{+}}\right)\Theta(-\alpha \beta)\right],
\end{eqnarray}
\end{widetext}
where $ z_{\pm}=(\beta/\alpha) \pm\sqrt{(\beta/\alpha)^2+\varepsilon} $ and $ \varepsilon =\tilde{\gamma}^2(E-E_{\rm v})/\alpha^3$ is energy deviation from VHS in dimensionless unit. The integration over $\bm p$  is extended to infinity. Here $ {\rm K}(z) $ is the complete elliptic integral of the first kind, $ {\rm sgn}(r)$ is the sign function defined as $ {\rm sgn}(r)=-1 $ for $r<0$ and $1$ for $ r\geq 0 $, and $ \Theta(r)\equiv\frac{1}{2}[1-{\rm sgn}(-r)] $ is the step function so that $\Theta(0)=0$.}

\begin{figure}
\includegraphics[width=0.5\textwidth]{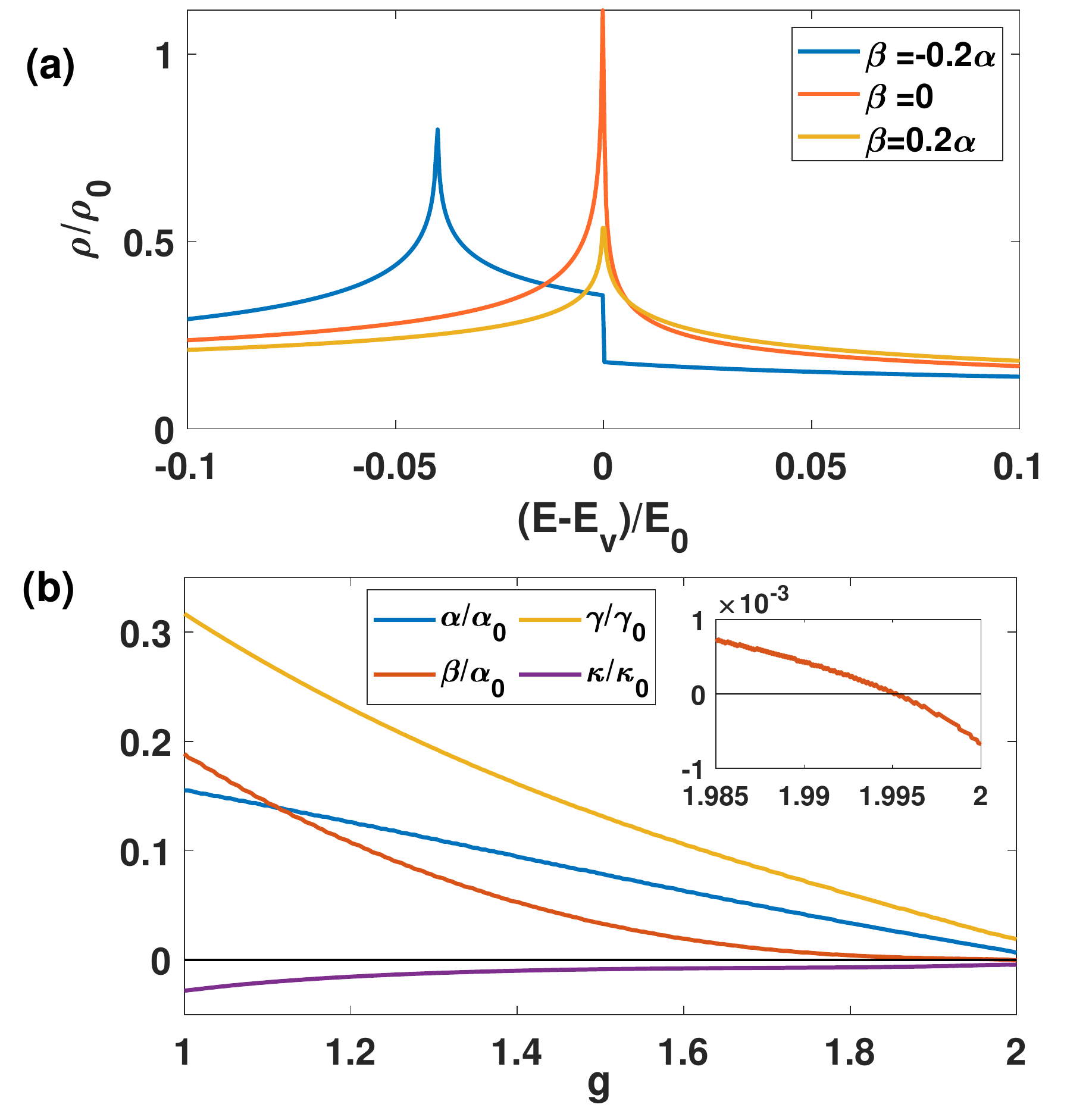}
\centering
\caption{{(a) DOS of Eq. (\ref{eq_DOS}) with different $\beta$, where $ E_{0}=\alpha^3/\tilde{\gamma}^2,\rho_{0}=\alpha^{-1} $. (b) Normalized coefficients of Eq. (\ref{expand}) as functions of $g$ when $ g'=1.2g $. Here $ \alpha_0=\lambda v,\gamma_0 =\lambda^2 v $ and $ \kappa_0 =\lambda^3 v $. Inset of (b) is zoom-in plot where $\beta$ changes sign near high-order VHS while $ \alpha,\gamma,\kappa $ do not.}}\label{fig_2}
\end{figure}

The DOS with different $\beta$ in Eq. (\ref{eq_DOS}) are shown in Fig. \ref{fig_2}a. In the following we will discuss the asymptotic behavior of the DOS near ordinary and high-order VHS. When $ \beta\neq 0 $, the DOS diverges logarithmically at ordinary VHS:
\begin{eqnarray}\label{oDOS}
\rho(E)
=\frac{1}{4\pi^2\sqrt{{\alpha |\beta |}}}\times
\begin{cases}
\log\frac{\Lambda}{|E-E_{\rm v}|} & \beta >0\\
{\sqrt{2}}\log\frac{\Lambda}{|E-E_{\rm v}-\delta |} & \beta <0\\
\end{cases}
\end{eqnarray}
where {$ \Lambda =64\alpha\beta^2/\tilde{\gamma}^2 $ is the {high}-energy cutoff}. In the limit $\beta\rightarrow 0$, the prefactor in Eq. (\ref{oDOS}) diverges as $1/\sqrt{|\beta|}$, indicating a strong increase of DOS as the VHS approaches a high-order one.
Right at the transition point $ \beta=0 $, we find power-law divergent DOS near high-order VHS
\begin{eqnarray}\label{DOS}
\rho(E)
&=&\frac{C}{\sqrt[4]{4\alpha\tilde{\gamma}^2}}\times
\begin{cases}
(E-E_{\rm v})^{-\frac{1}{4}} & E>E_{\rm v}\\
{\sqrt{2}}(E_{\rm v}-E)^{-\frac{1}{4}} & E<E_{\rm v}
\end{cases}
\end{eqnarray}
where $ C=(2\pi)^{-\frac{5}{2}}\Gamma(\frac{1}{4})^2=0.133 $.
This power-law divergence with exponent $-1/4$ is stronger than the logarithmic divergence at ordinary VHS. Another key difference is that the diverging part of DOS around this high-order VHS is inherently asymmetric with respect to $E_{\rm v} $, as reflected by the $\sqrt{2}$ factor in Eq. (\ref{DOS}). This is in contrast with ordinary VHS where the DOS above and below are asymptotically symmetric.
{Using Eq. (\ref{eq_DOS}), we can fit the experimental data of tunneling conductance in Ref. \cite{Abhay} with finite broadening, as shown in Fig. \ref{fig_0}a and Supplemental Material.}

At the high-order VHS {($\beta=0$)}, the dispersion has the following scaling invariance
\begin{eqnarray}\label{eq_sc}
E(\lambda^{\frac{1}{2}}p_{x},\lambda^{\frac{1}{4}}p_{y})=\lambda E(p_{x},p_{y})
\end{eqnarray}
with arbitrary $ \lambda>0 $. Here we set $ E_{\rm v}=0 $ for simplicity. Including higher order terms or the third and fourth order terms other than $ \gamma,\kappa $ in Taylor expansion (\ref{expand}) will break scaling invariance (\ref{eq_sc}).
Notice that the scaling dimensions along $ p_x $ and $ p_y $ directions are different $ [p_{x}]=\frac{1}{2},[p_{y}]=\frac{1}{4} $, with scaling dimension of energy $[E]=1$.
From the scaling invariance (\ref{eq_sc}) we immediately obtain the scaling dimension of DOS as $ [\rho]=[p_x]+[p_y]-[E]=-\frac{1}{4} $, the same as Eq. (\ref{DOS}). To directly reveal the power-law behavior of experimental data, in Fig. \ref{fig_0}b we plot energy deviation $ |E-E_{\rm v}| $ and experimental DOS including background contribution both in logarithmic scale. In the log-log plot, two sides of the high-order VHS peak form two parallel lines with the same slope $ -1/4 $, and the particle-hole asymmetry ratio is 1.32 close to $ \sqrt{2} $, which are both consistent with Eq. (\ref{DOS}).

Besides $\alpha$ and $\beta$, the Fermi contours in momentum space depend on both $ \gamma $ and {$\kappa$}, while the DOS in energy domain depends {only on} $ \tilde{\gamma} $. This is {because the same scaling dimension of $p_x$ and $p_y^2$ allows} the nonlinear transform
\begin{eqnarray}\label{eq_trans}
\tilde{p}_{x}=p_{x}-\frac{\gamma}{2\alpha}p_{y}^2,\quad \tilde{p}_{y}=p_{y},
\end{eqnarray}
under which the dispersion (\ref{expand}) becomes $ E-E_{\rm v}=-\alpha \tilde{p}_{x}^2+\beta \tilde{p}_{y}^2+\tilde{\gamma}^2\tilde{p}_{y}^4/(4\alpha) $. The nonlinear transform (\ref{eq_trans}) changes the shape of Fermi contours while preserves area element and hence leaves DOS in Eq. (\ref{eq_DOS}) invariant.

\begin{figure}[ht]
\includegraphics[width=0.5\textwidth]{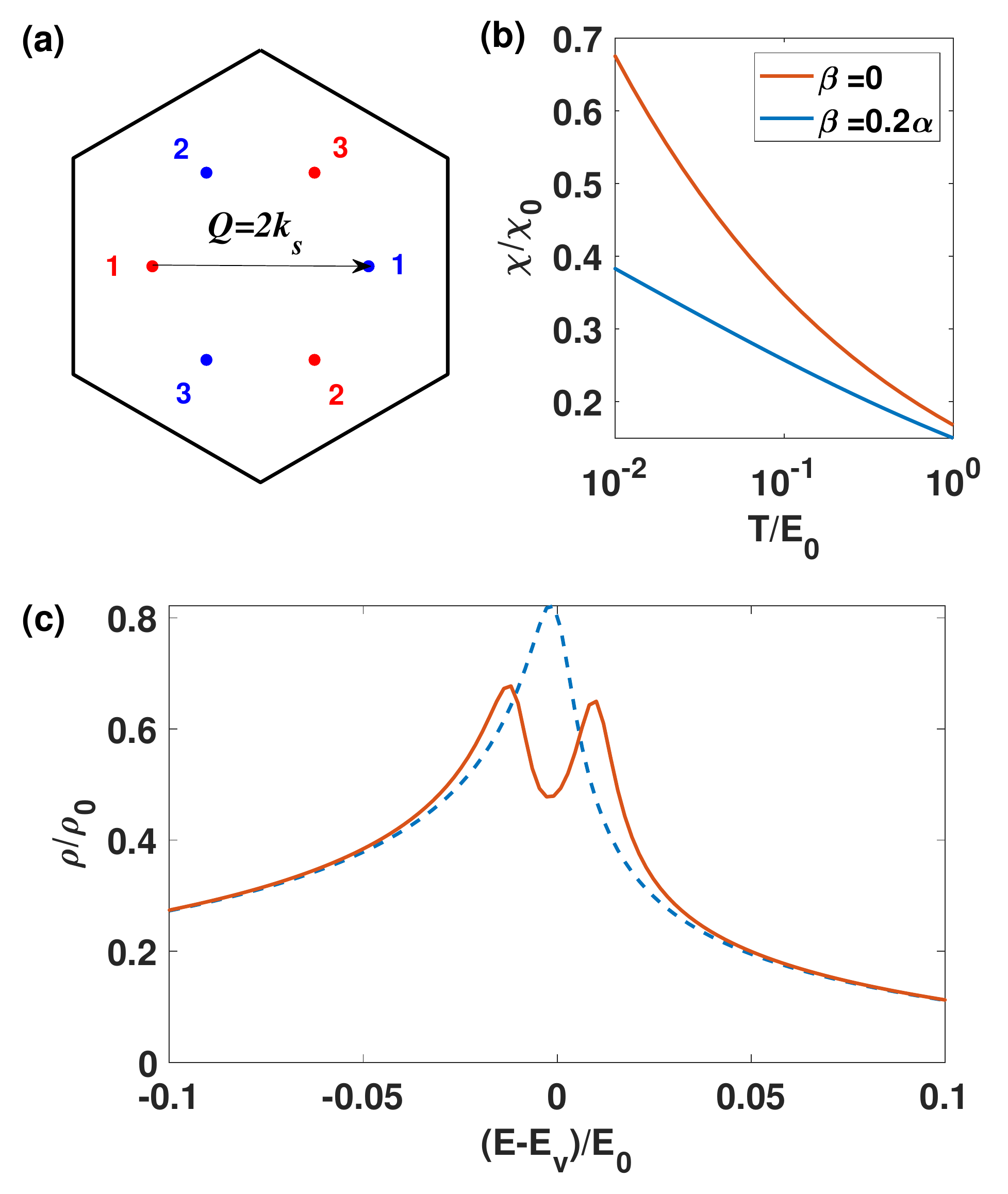}
\centering
\caption{{(a) Six high-order VHS in MBZ, where different colors denote different valleys, and 1, 2, 3 denote three $\Gamma M$ directions. (b) Temperature dependence of intervalley susceptibility $\chi\equiv\chi(\bm{Q},E_\text{v})$ at ordinary ($\beta =0.2\alpha$) and high-order ($\beta =0$) VHS with $\bm Q=2\bm k_s$ and $ \chi_{0}=\alpha^{-1} $. (c) Density of states with (solid) and without (dashed) intervalley density wave when chemical potential is at high-order VHS on the hole side $E_{\rm v}<0$. Here $ E_{0}=\alpha^3/\tilde{\gamma}^2 $ and details can be found in the Supplemental Material.}}\label{fig_3}
\end{figure}

We have thus established by continuity argument the existence of high-order VHS in the moir\'e band of TBG at a {certain} critical {twist} angle $\theta_c$. 
The value of $\theta_c$ depends on the model we employ. As an illustrative example, we consider the continuum model \cite{Neto,Neto2,MacDonald,KoshinoFu} of TBG, where approximate particle-hole symmetry holds, and the Fermi surface topology is solely determined by two dimensionless parameters $ g=\lambda u/v,g'=\lambda u'/v $. Here $u$ and $u'$ denote interlayer hoppings, $v$ is the monolayer Dirac velocity and $\lambda$ is the moir\'e wavelength. We further fix the ratio $ g'/g=u'/u=1.2 $ as calculated for corrugated structure at $\theta \sim 1^\circ$ \cite{KoshinoFu}. At given $g\in[1,2]$, we calculate the Taylor coefficients $\alpha,\beta,\gamma,\kappa $ by numerical derivatives of energy with respect to momentum at VHS on the hole side, and the result is shown in Fig. \ref{fig_2}b. It can be found that $ \gamma^2+4\alpha\kappa $ is always positive when $ g\in [1,2] $, and across $ g_{c}\approx 1.995 $, $ \beta $ changes sign {(Fig. \ref{fig_2}b inset)} while $ \alpha,\gamma,\kappa $ do not. With appropriate and reasonable values of $u$ and $v$, from $ g_c $ we can obtain the critical twist angle $ \theta_{c}\approx 1^\circ $, consistent with experiments. Details of continuum model can be found in the Supplemental Material. We may include more ingredients such as lattice relaxation \cite{NamKoshino,Stephen} and strain \cite{BiZhen} to better describe TBG, thus the Fermi surface topology will depend on more details of the model. Nevertheless, the existence of high-order VHS and critical {twist} angle $\theta_c$ is robust \cite{BiZhen}.

{{\it Many-body phenomena near high-order VHS}.}
When chemical potential is near high-order VHS, many-body effects can be drastic as the strongly diverging DOS leads to divergences in noninteracting susceptibilities of various channels. This signals a strong tendency to various broken symmetry states around van Hove filling, when electron-electron interaction is taken into account. Indeed, the recent STS measurements found that when the Fermi energy approaches the van Hove energy under doping, the VHS peak in DOS splits into two new ones \cite{Abhay, Stevan}.

A detailed analysis of interacting electrons near high-order VHS is beyond the scope of this work. Instead we discuss two possible scenarios for the observed splitting of the VHS peak {near the Fermi energy}.

First, strains in experimental samples can split DOS peak by breaking rotation symmetries which relate VHS along different directions. Though the energy splitting due to strain may be small at single-particle level, Coulomb interaction $U$ can give rise to the Stoner-type enhancement factor $(1-U\chi_s)^{-1}$ within the random phase approximation (RPA). Since the noninteracting susceptibility $ \chi_s $ {reflects the divergent DOS}, strain effect on high-order VHS can be greatly enhanced by interaction.

Second, an intervalley density wave order can split a DOS peak by spontaneously breaking translation symmetry. To show such a density wave instability, we calculate intervalley susceptibility $\chi(\bm{q})$ with finite momentum $\bm q$. There are in total six high-order VHS
from two valleys, located at three $\Gamma M$ lines {(Fig. \ref{fig_3}a)}.
When we consider two high-order VHS along the same $\Gamma M$ line but from opposite valleys, $\chi(\bm{q})$ is sharply enhanced near wave vector $ \bm q=\bm Q $ due to the power-law divergent DOS from both valleys, where $\bm{Q}=2\bm{k}_s$ is the momentum separation between the two VHS (Fig. \ref{fig_3}a).
As a result, when approaching low temperatures, the intervalley susceptibility $\chi(\bm{Q})$ of high-order VHS diverges more rapidly than logarithm (Fig.~\ref{fig_3}b), indicating a stronger {tendency toward} density wave instability than one with ordinary VHS.
The DOS peak at VHS energy $ E_{\rm v} $ is split into two in the presence of intervalley density wave (Fig. \ref{fig_3}c).

To summarize, we propose that the proximity to high-order VHS underlies correlated electron phenomena in TBG near magic angle. We reveal a distinctive feature of high-order VHS---power-law divergent DOS with an asymmetric peak, which compares well with the recent STS data on magic-angle TBG. We also discuss nematic and density wave instabilities due to electron-electron interaction near van Hove filling, and {illustrate} the splitting of VHS in these broken symmetry states.

It is worth examining the similarity and difference between the two scenarios favoring correlated electron phenomena in TBG: flat band \cite{MacDonald} and (high-order) VHS. Both scenarios rely on large DOS enhancing electron correlation. Obviously, the largest possible DOS is realized in the limit of a completely flat band \cite{Vishwanath}. However, a completely flat band is  difficult to achieve under realistic conditions. On the other hand, the VHS scenario is more likely to be relevant when the bandwidth is not so small in comparison to electron interaction. In this scenario, the DOS near Fermi energy matters more than those far away, and it is further increased by proximity to high-order VHS. Importantly, to achieve high-order VHS generally requires tuning the band structure with only a single parameter, whereas to achieve a completely or extremely flat band usually requires more fine tuning.

Broadly speaking, moir\'e superlattices in 2D materials offer an unprecedented platform for studying correlated electron phenomena near VHS. Electrostatic gating enables doping these systems to van Hove filling without introducing disorder. The great tunability of moir\'e band structure by twist angle, electric field, strain and other means grants access to high-order VHS, where the strongly divergent DOS promises a plethora of many-body phenomena. Such examples include TBG discussed in the maintext and trilayer graphene on boron nitride discussed in Supplemental Material. The time is coming for {\it creating} strong electron correlation by {\it designing} single-particle dispersion around VHS.

\textit{Acknowledgment.} We thank Alex Kerelsky and Abhay Narayan Pasupathy for sharing experimental data with us. This work is supported by DOE Office of Basic Energy Sciences, Division of Materials Sciences and Engineering under Award DE-SC0010526. LF is partly supported by the David and Lucile Packard Foundation.

\appendix
\setcounter{figure}{0}
\renewcommand{\thefigure}{S\arabic{figure}}
\setcounter{equation}{0}
\renewcommand{\theequation}{S\arabic{equation}}

\begin{widetext}
\section*{Supplemental Material for ``Magic of high-order van Hove singularity"}
\subsection{Theoretical fitting of tunneling conductance in twisted bilayer graphene}
\subsubsection{Model-dependent fitting}
The dispersion near a van Hove singularity (VHS) is
\begin{eqnarray}\label{seq_E}
E-E_{\rm v}=-\alpha p_{x}^2+\beta p_{y}^2 +\gamma p_{x}p_{y}^2+\kappa p_{y}^4,
\end{eqnarray}
and the corresponding density of states (DOS) is
\begin{eqnarray}\label{seq_DOS}
\rho(E) =\frac{1}{\sqrt{2}\alpha\pi^{2}}Q(\varepsilon ,r),\quad \varepsilon =\frac{\tilde{\gamma}^2}{\alpha^3}(E+i\eta -E_{\rm v}),\quad r=\frac{\beta}{\alpha},
\end{eqnarray}
where $ Q\equiv{\rm sgn}(r)\left[{\rm Re}f+\Theta(-r){\rm Im}g\right] $, $ \tilde{\gamma}=\sqrt{\gamma^2+4\alpha\kappa} $ and
\begin{eqnarray}
f(\varepsilon ,r)=\frac{1}{\sqrt{z_{-}}}{\rm K}\left(1-\frac{z_{+}}{z_{-}}\right),\quad
g(\varepsilon ,r)=\frac{2}{\sqrt{z_{+}}}{\rm K}\left(\frac{z_{-}}{z_{+}}\right)
\end{eqnarray}
with $ z_{\pm}=r\pm\sqrt{r^2+\varepsilon} $. The elliptic integral we use is defined as
\begin{eqnarray}
{\rm K}(z)=\int_{0}^{\pi/2}\frac{d\theta}{\sqrt{1-z\sin^2\theta}}.
\end{eqnarray}

In Eq. (\ref{seq_DOS}) we assume the dispersion (\ref{seq_E}) extends to the whole momentum space, while for realistic systems Eq. (\ref{seq_E}) holds only for a finite range $ |E-E_{\rm v}|<\Omega $. To account for the high-energy cutoff $ \Omega $, we introduce a negative background to the DOS expression (\ref{seq_DOS}), and hence tunneling conductance becomes
\begin{eqnarray}\label{seq_G}
G(E) =G_{0}Q(\varepsilon ,r)-G_{c},
\end{eqnarray}
with two additional parameters $ G_{0} $ and $ G_{c} $. Here $ G_0 $ is due to the tunneling matrix element between the sample and the tip in STS experiments, and $G_c$ is due to the high-energy cutoff $\Omega$. Notice that Eq. (\ref{seq_G}) only applies to energy range $ |E-E_{\rm v}|<\Omega $.

In total we have 6 parameters $ G_0,G_c,E_{\rm v},\eta,r,E_{0}\equiv\alpha^3/\tilde{\gamma}^2 $ in the fitting of tunneling conductance. However, due to the scaling property $Q(\lambda^2\varepsilon ,\lambda r)=\lambda^{-\frac{1}{2}} Q(\varepsilon ,r)$, there are only 5 independent parameters $ G_0,G_c,E_{\rm v},\eta $ and $ w\equiv E_{0}r^2{\rm sgn}(r)=\alpha\beta^2{\rm sgn}(\beta)/\tilde{\gamma}^2 $. Here $ |w| $ is the low-energy cutoff of the high-order VHS, namely in energy range $ |w|<|E-E_{\rm v}|<\Omega $ the VHS can be treated approximately high-order. When $ w=0 $ the VHS is exactly high-order. 

The optimal fitting parameters in least squares fitting are given below in Table. \ref{S} for different samples, and fitting results are shown in Fig. \ref{sfig_1}.
We can treat $ w $ as the single indicator of the topological transition of VHS similar to $\beta$ in Eq. (\ref{seq_E}). From Table. \ref{S}, it can be found 
\begin{eqnarray}
w(2.02^\circ)\gg w(1.10^\circ)\approx 0>w(0.79^\circ),
\end{eqnarray}
indicating a topological transition of VHS.

\begin{table}
\centering
\begin{tabular}{p{1cm}|p{2cm}p{2cm}p{2cm}p{2cm}p{2cm}|p{1cm}} \hline
$\theta$ & $G_{0}$ (nS) & $G_c$ (nS) & $E_{\rm v}$ (meV) & $\eta$ (meV) & $w$ (meV) & band \\
\hline\hline
2.02$^\circ$ & 6.112$ G_{\rm m} $ & 3.108$ G_{\rm m} $ & -94.15 & 23.29 & 49.69 & $-$\\
            \hline
1.10$^\circ$ & 98.22 & 67.57 & -36.05 & 2.786 & 3.360 & $-$\\
             & 76.34 & 57.6 & 16.72 & 3.556 & -0.1442 & $+$\\
             \hline
0.79$^\circ$ & 30.89 & 16.22 & -3.299 & 1.822 & -4.271 & $-$\\
             & 32.82 & 16.60 & 0.7944 & 2.074 & -5.8787 & $+$\\
\hline
\end{tabular}
\caption{Fitting parameters of tunneling conductance data at twist angle $ \theta $. In the last column $ \pm $ denote positive/negative energy band respectively. For $\theta =2.02^\circ$, $G_{\rm m}$ denotes the maximal value of tunneling conductance data.}
\label{S}
\end{table}

\begin{figure}
\centering
\subfloat[$\theta=$ 2.02$^\circ$.]{{\includegraphics[width=12cm]{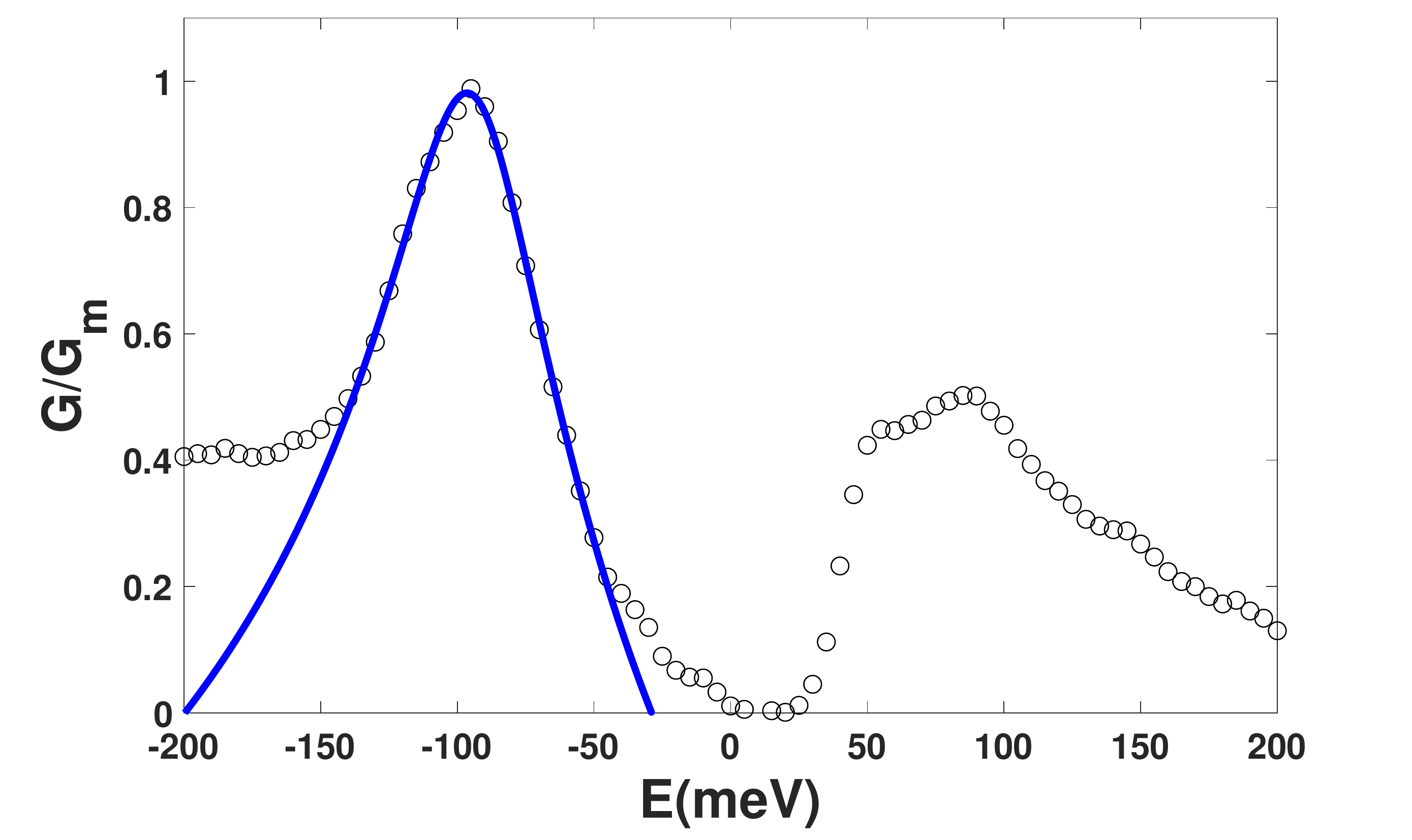} }}
\qquad
\subfloat[$\theta=$ 1.10$^\circ$.]{{\includegraphics[width=12cm]{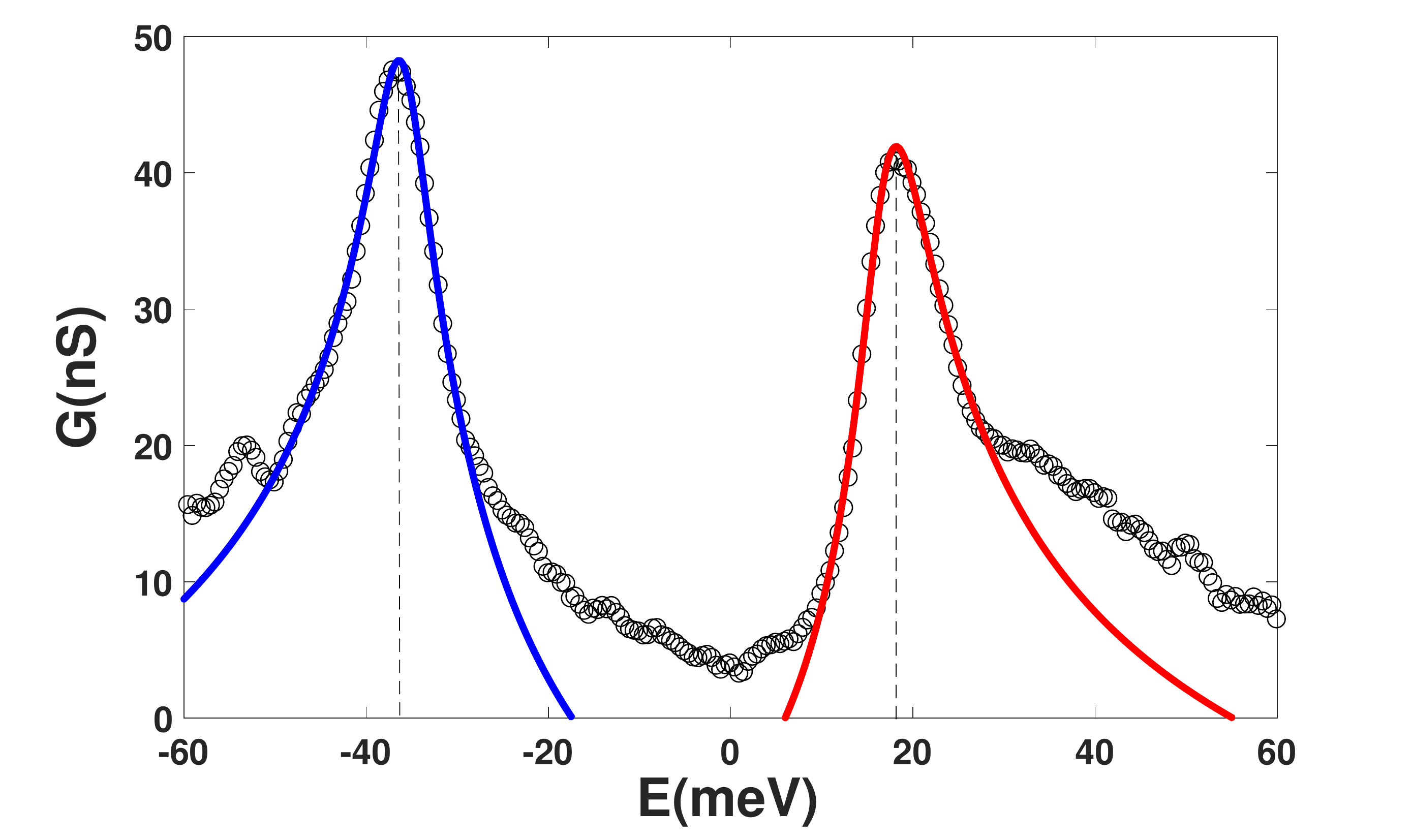} }}
\qquad
\subfloat[$\theta=$ 0.79$^\circ$.]{{\includegraphics[width=12cm]{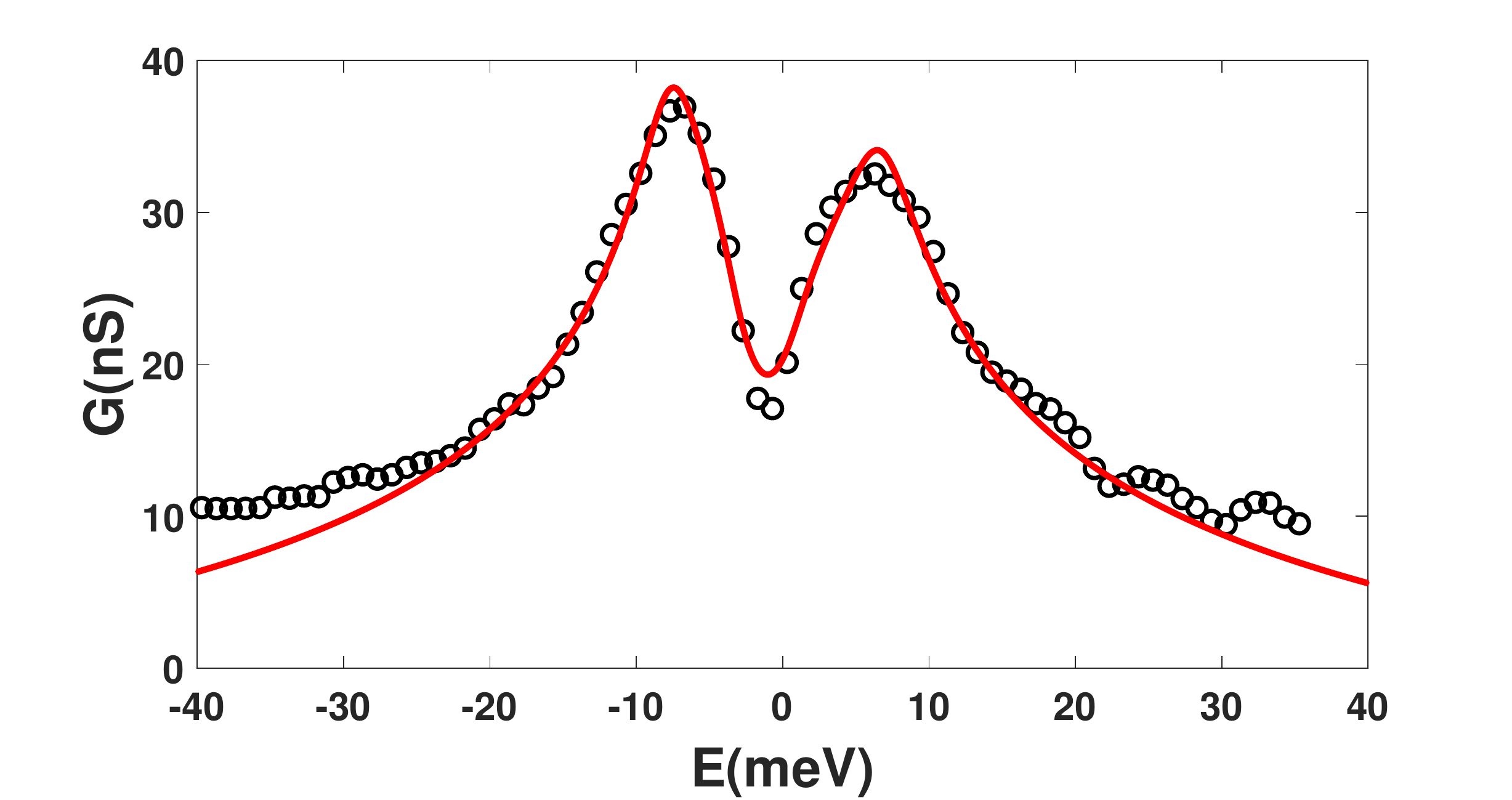} }}%
\caption{Open circles are tunneling conductance data $ G $ of twisted bilayer graphene \cite{Abhay}. Solid lines are Eq. (\ref{eq_G}) with parameters in Table. \ref{S}. In (a), $G_{\rm m}$ denotes the maximal value of tunneling conductance data. Dashed lines in (b) denote positions of van Hove energy, indicating the asymmetry of van Hove peaks.}
\label{sfig_1}
\end{figure}

\subsubsection{Model-independent fitting}
There are two distinct types of divergent DOS and hence conductance peaks in this work and also in experimental data. The first type is due to ordinary VHS, which is symmetric and logarithmic, and another type is due to high-order VHS, which is intrinsically asymmetric and power-law. These two types of conductance peaks follow different functional forms as follows ($o$ denotes ordinary VHS and $h$ denotes high-order VHS)
\begin{eqnarray}
G_{o}(E)=A\log{|E-E_{\rm v}|}-G_c,\quad
G_{h}(E)=A{\rm Re}(E_{\rm v}-E)^{-1/4}-G_c,
\end{eqnarray}
each with three parameters $A,E_{\rm v}$ and $G_c$. Among them, $E_{\rm v}$ denotes the energy of conductance peak, and $G_c$ is due to high-energy cutoff. 

In Fig. \ref{sfig_2}, we plot conductance data $G$ near the peaks at two different angles $ \theta =2.02^\circ $ and $ 1.10^\circ $ respectively, where energy is plot in logarithmic scale. When $ \theta =2.02^\circ $, the conductance is plot in original scale. Since the VHS is ordinary, conductance peak is symmetric and logarithmic, two sides of the peak will collapse into the same line. When $ \theta =1.10^\circ $, the conductance is plot in logarithmic scale. Since the VHS is high-order, conductance peak is asymmetric and power-law, two sides of the peak will follow two parallel lines with the same slope -1/4 but different vertical intercepts. 

\begin{figure}
\centering
{{\includegraphics[width=16cm]{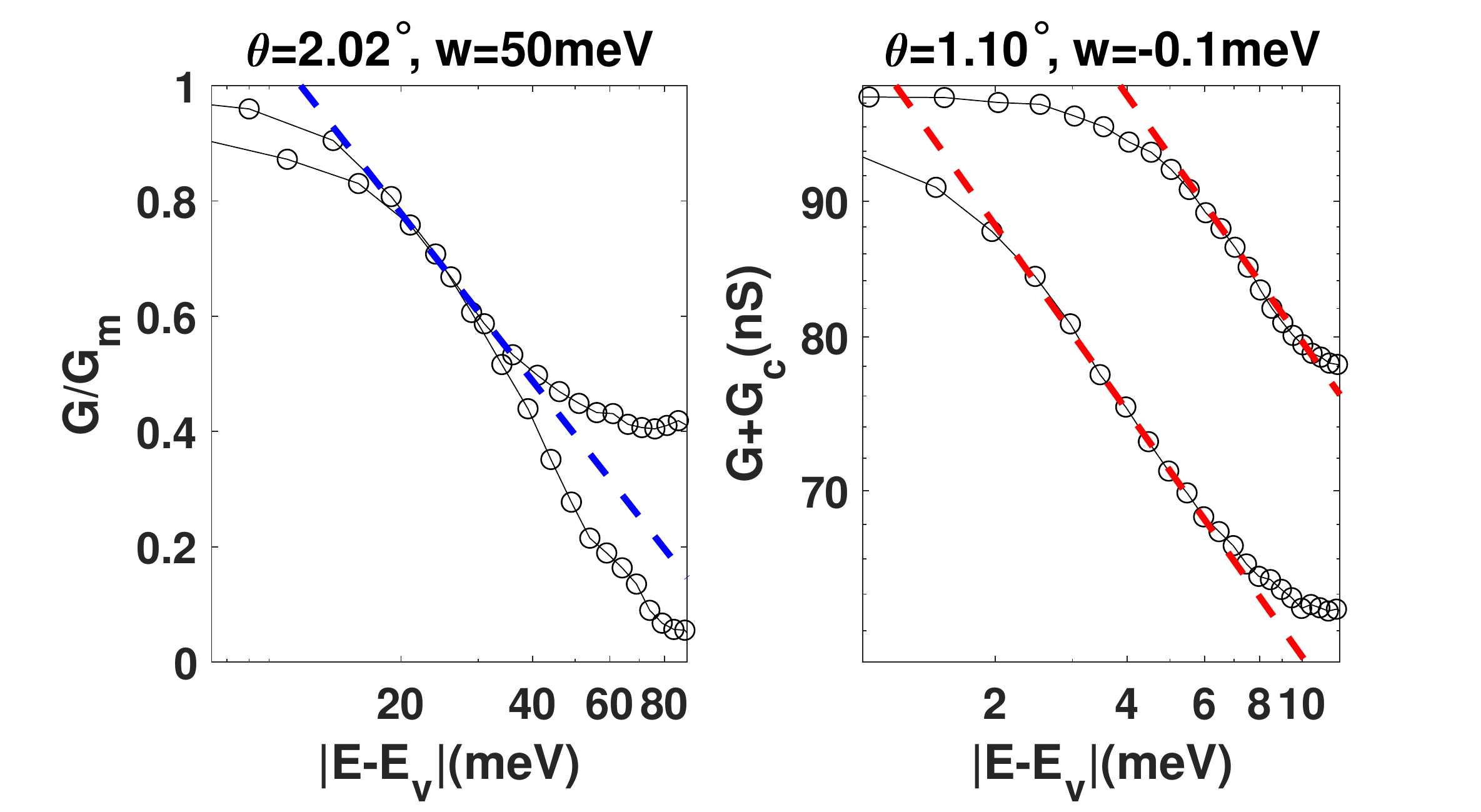} }}
\caption{Conductance peaks induced by ordinary (left panel) and high-order (right panel) VHS. In both figures, energy is measured from conductance peak energy $E_{\rm v}$ and plot in logarithmic scale, open circles are tunneling conductance data $ G $, and two lines denote two sides of the conductance peak. In the left panel $ \theta=2.02^\circ $, the conductance peak is at negative energy $ E_{\rm v}=-94.15{\rm meV}<0 $, and two sides of the peak collapse to the same dashed blue line when $\eta<|E-E_{\rm v}|<|w|$. In the right panel $ \theta=1.10^\circ $, the conductance peak is at positive energy $ E_{\rm v}=16.72{\rm meV}>0 $, conductance is in logarithmic scale and two sides of the peak follow two parallel lines when $|E-E_{\rm v}|\gg|w|$. The two dashed red lines have the same slope $-1/4$, and the difference between vertical intercepts is 0.28, deviating from the theoretical value $ \log\sqrt{2}=0.35 $ caused by imperfect high-order VHS and finite broadening.}
\label{sfig_2}
\end{figure}

\subsection{Continuum Model}
To determine the value of $\theta_c$, we calculate the  moir\'e band structure and track its evolution with twist angle. It is important to note that our calculation of TBG band structure is only to illustrate the Lifshitz transition of VHS and existence of high-order VHS in TBG, thus many ingredients such as strain and in-plane lattice relaxation are neglected.  

Our calculation uses a generalized  continuum model \cite{KoshinoFu}
\begin{eqnarray}\label{fcm}
H(\bm k,\bm r)=
\begin{pmatrix}
vR_{\theta/2}\bm p_{1}\cdot\bm\sigma & U(\bm r)\\
U^{\dagger}(\bm r) & vR_{-\theta/2}\bm p_2\cdot\bm\sigma
\end{pmatrix},\quad
U(\bm r)=
\sum_{j=0}^{2}
\begin{pmatrix}
u & u'\omega^{-j}\\
u'\omega^{j} & u
\end{pmatrix}
e^{i\bm G_{j}\cdot\bm r},
\end{eqnarray}
with $\omega =e^{2\pi i/3}$.
Here $\bm \sigma$  acts on the two components of massless Dirac fermion at $\bm K$ point of each layer. Due to the relative orientation of the two layers, the Dirac spinors on the two layers are rotated by $\pm \theta/2$ respectively under the orthogonal rotation matrix $R_{\pm\theta/2} $. $ \bm p_{l}=\bm k-\bm K_{l} $ denotes momentum deviation from the $\bm K$ point in layer $l=1,2$. The $2\times 2$ matrix $U(\bm r)$ denotes interlayer tunneling and is periodic in real space $ U(\bm r+\lambda\bm e_{\pm})=U(\bm r) $, where $ \lambda\bm e_{\pm} $ are two primitive vectors of TBG moir\'e superlattice, and $ \lambda =a/(2\sin\frac{\theta}{2}) $ is the moir\'e wavelength at twist angle $\theta$ with graphene lattice constant $a$. $ U(\bm r) $ contains two independent parameters $ u $ and $ u' $ associated with interlayer tunneling within the same sublattice and between the two sublattices respectively. Due to the out-of-plane lattice relaxation, AA regions have larger interlayer distance than AB regions \cite{Abhay}, resulting in $ u' < u$ in improved continuum model of TBG band structure \cite{KoshinoFu}.

At small $\theta$, Fermi surface topology and hence VHS in the continuum model (\ref{fcm}) are solely determined by two dimensionless parameters $ g\equiv \lambda u/v $ and $ g'\equiv \lambda u'/v $ {,where $v$ is the Dirac velocity and $\lambda$ is the moir\'e wavelength}. 
This can be seen by following scaling relation
\begin{eqnarray}
H(\lambda^{-1}\bm k,\lambda\bm r)=\frac{v}{\lambda}
\begin{pmatrix}
\bm p_{1}\cdot\bm\sigma & f(\bm r)\\
f^{\dagger}(\bm r) & \bm p_2\cdot\bm\sigma
\end{pmatrix},\quad
f(\bm r)=
\sum_{j=0}^{2}
\begin{pmatrix}
g & g'\omega^{-j}\\
g'\omega^{j} & g
\end{pmatrix}
e^{i\bm n_{j}\cdot\bm r},\quad
\bm n_{0}=\bm 0,\quad\bm n_{1,2}=(\mp\frac{1}{2},\frac{\sqrt{3}}{2}).
\end{eqnarray}

Dimensionless parameters $g, g'$ increase with the moir\'e wavelength as $\theta$ decreases; they also increase with interlayer tunneling under pressure.
For simplicity, in the following we fix the ratio $ g'/g=u'/u=1.2 $ as calculated for relaxed structure at $\theta \sim 1^\circ$ \cite{KoshinoFu} and study VHS evolution with the single parameter $ g $.
Since electron and hole sides are approximately symmetric, we focus on VHS in the hole side.

At $g=1$ there are three ordinary VHS on $\Gamma M$ (Fig. 2a of maintext), while at $ g=2 $ there are six ordinary VHS on two sides of $\Gamma M$ (Fig. 2c of maintext). According to our analysis above, this implies a topological transition of VHS occurs at $ g=g_{c}\in (1,2) $.
To determine the value $g_c$, we calculate the Taylor coefficients $\alpha,\beta,\gamma,\kappa $ by numerical derivatives of energy dispersion with respect to momentum components $ k_x,k_y $ at VHS point. As shown in Fig. 3b of maintext, $\alpha,\beta,\gamma,|\kappa| $ all decrease as $g$ increases, and across $ g_{c}\approx 1.995 $, $ \beta $ changes sign {(Fig. 3b inset)} while $ \alpha,\gamma $ stay positive and $ \kappa $ stays negative. It can be verified that through the whole process $ \gamma^2+4\alpha\kappa $ is positive so that the asymptotic behavior is always saddle point.

At $g<g_c$, we find that over an extended parameter range $ g\in [1.6, 1.8] $, the van Hove filling is close to two electrons per unit cell (where correlated electron phenomena are observed), 
in agreement with the observed sign change of Hall coefficient under doping and pressure \cite{Dean}. Importantly, in this range $ 0<\beta\ll\alpha $ and VHS can still be treated approximately as high-order.
At small $g-g_c>0$, the bandwidth decreases rapidly; the band structure calculated from continuum model becomes rather complex {above} $g_c$ \cite{Kasra}. It is possible that additional effects not captured in Eq. (\ref{fcm}), such as in-plane lattice distortion, become important in this regime \cite{NamKoshino,Stephen,BiZhen}. Nonetheless, the existence of topological transition of VHS as a function of $\theta$ is a robust feature.

To translate the parameter $g$ into the actual twist angle requires the knowledge of Dirac velocity $v$ and interlayer hoppings $u, u'$, which are subject to some uncertainty. It is known that Coulomb interaction increases the Dirac velocity substantially at low energy \cite{Geim,Elias,Schliemann}. It is recently proposed that this increased velocity  may account for the discrepancy between the  moir\'e bandwidth found in STS measurements and in previous calculations \cite{Abhay}.
When we use an increased Dirac velocity $ va^{-1}=2.41 $eV, 13\% larger than DFT value and adopt the interlayer hopping parameters $ u=79.7 $meV, $ u'=97.5 $meV from Ref. \cite{KoshinoFu},  we find $\theta_c =0.95^\circ $,
and the experimentally established range of magic angle $ \theta_{\rm exp}\in[1^\circ, 1.2^\circ] $ corresponds to the range $ g_{\rm exp}\in [1.6,1.9] $ where $ 0<\beta\ll\alpha $ and VHS is approximately high-order.
We can also apply pressure to change interlayer couplings $u,u'$ and hence $ g,g' $. In this way, high-order VHS can be achieved at larger twist angles.
\subsubsection{Total density of states from continuum model}
We can also use continuum model instaed of effective model (\ref{seq_E}) to compute the total density of states. In Fig. \ref{fig_tdos} we plot the Fermi contours and corresponding total density of states near van Hove singularity in continuum model with different coupling constant $g$.

\begin{figure*}
\centering
\includegraphics[width=1.05\hsize]{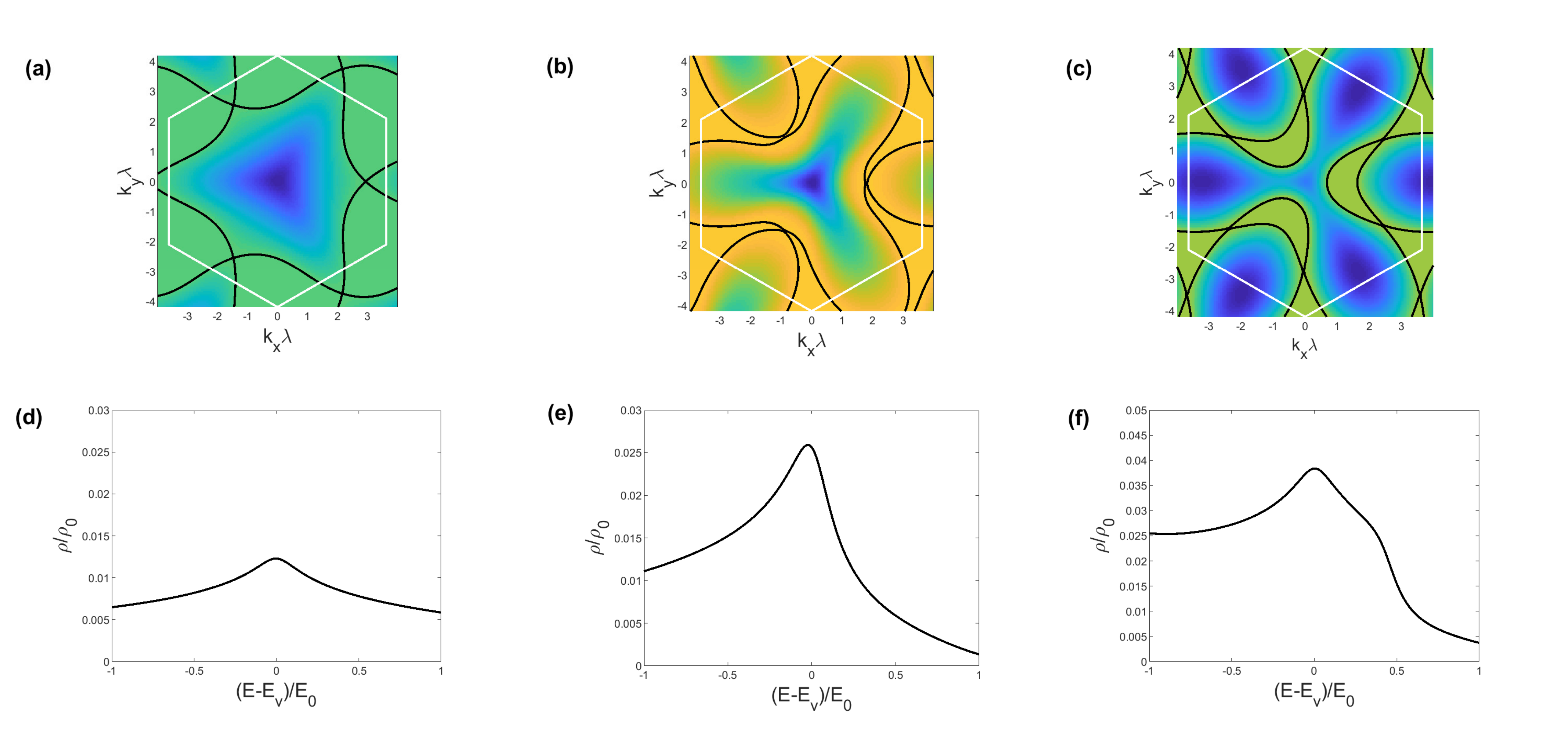}
\caption{
(a,b,c) Energy contours and (d,e,f) total density of states calculated from continuum model. The coupling constants are $ g'=1.2g $ and (a,d) $ g=1 $, (b,e) $ g=1.9 $ and (c,f) $ g=2 $. Here $ E_{\rm v} $ is the energy of VHS, $ E_{0}=\alpha^3/\tilde{\gamma}^2,\rho_{0}=\alpha^{-1} $, and $ \alpha,\beta,\tilde{\gamma} $ are calculated from numerical derivatives of band structure at VHS. The broadening in DOS calculation is $ \eta =2.2\times 10^{-3}E_{0} $.
}
\label{fig_tdos}
\end{figure*}

\subsubsection{Local density of states from continuum model}

In the maintext and also previous sections, we calculate the total density of states which is the averaged DOS over the whole sample. In this section we show that at magic angle, even local density of states (LDOS) also exhibits behaviors of high-order VHS, both in AA and AB regions.

In Fig. \ref{fig_ldos}, by continuum model we calculate LDOS in AA and AB regions of TBG at magic angle as functions of energy, where both LDOS and energy are plot in log scale and $ E_{\rm v} $ denotes the energy of conductance peak. Red and blue colors denote peaks at positive and negative energy sides respectively. We find that LDOS in both AA and AB regions show asymmetric power-law divergent peaks, while LDOS in AA region is much larger than LDOS in AB region, which is consistent with experimental data in arXiv:1812.08776 by A. Kerelsky \textit{et. al}.

\begin{figure}
\centering
\includegraphics[width=1\hsize]{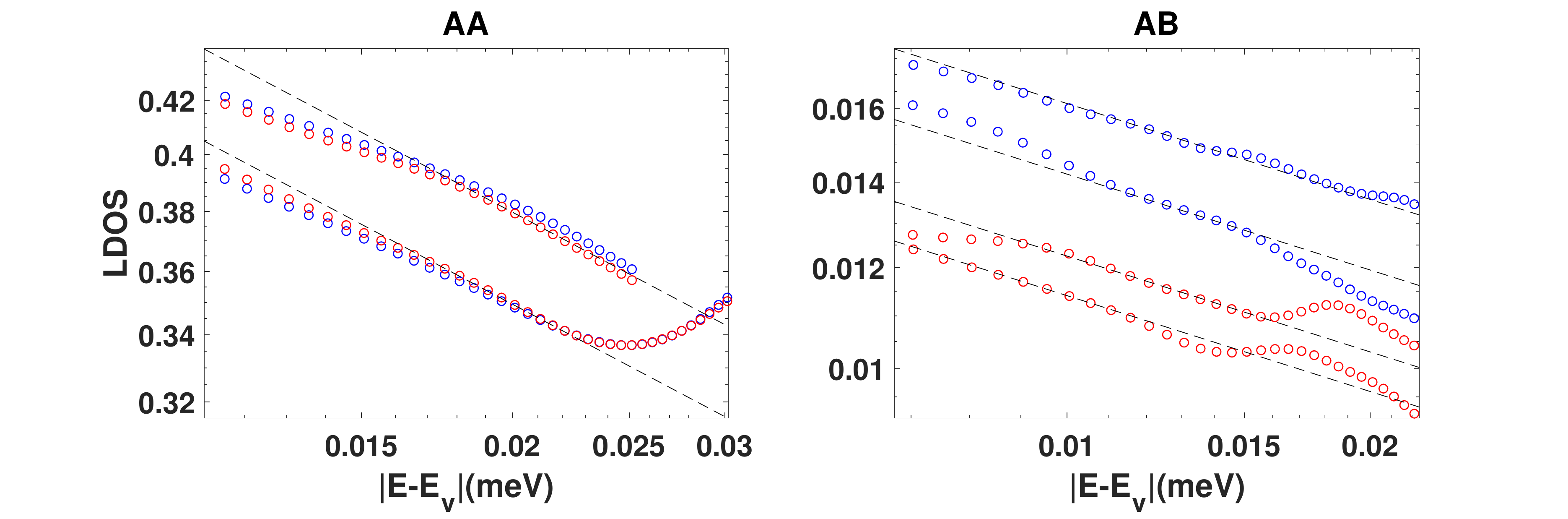}
\caption{
Local density of states (LDOS) in AA and AB regions of twisted bilayer graphene at magic angle, where both LDOS and energy are plot in log scale.
}
\label{fig_ldos}
\end{figure}

\subsection{Mean-field Hamiltonian of interacting high-order VHS}
With finite intervalley density wave order parameter $ \Delta $, the low-energy Hamiltonian of two high-order VHS on the same $\Gamma M$ line is
\begin{eqnarray}
\mathcal{H}(\bm p)=E_{\rm v}-\alpha p_{x}^2+\kappa p_{y}^4+\gamma p_{x}p_{y}^2\tau_z +\Delta\tau_x ,
\end{eqnarray}
where Pauli matrices $\bm{\tau}$ act on the valley indices.
From this Hamiltonian Fig. 4c of maintext is plotted, and the order parameter is $ \Delta =0.01E_0 $ with broadening $ \eta =5\times 10^{-3}E_0 $.

\subsection{Trilayer graphene on boron nitride}

The low-energy physics of ABC-stacked trilayer graphene (ABC-TLG) is described by states at $A_1$ and $B_3$ sites near $ \pm\bm K $ points of Brillouin zone shown in Fig. \ref{fig1}c. The point group at $\bm K$ point is $ D_{3} $ with two generators $ C_{3z} $ and $ C_{2x} $. Since $A_1$ and $B_3$ are at rotation center $O$ and equivalent points, Bloch waves with momentum $ \bm K $ at $A_1$ and $B_3$ will carry zero angular momentum under $C_{3z}$ and furnish 1D representations $ A_1 $ and $A_2$ of $ D_{3} $. As a result, the band spectrum at $\bm K$ point is gapped. In the basis of $A_1$ and $B_3$, the 2 by 2 effective $ k\cdot p $ Hamiltonian near $+\bm K$ point reads
\begin{eqnarray}
H(\bm p)=-\mu +ap^2 +(\gamma +bp^2)\sigma_{x}+c
\begin{pmatrix}
0&p_{-}^3\\
p_{+}^3&0
\end{pmatrix}
\end{eqnarray}
up to the third order in $\bm p$, where $ p_{\pm}=p_{x}+ip_{y} $ and $ p=|\bm p|=|p_{\pm}| $. Here the parameters are (unit: eV)
\begin{eqnarray}
\mu =0.0027,\quad a=3.2708,\quad\gamma =-0.0083,\quad b=3.2165,\quad c=91.1443.
\end{eqnarray}

\begin{figure}
\centering
\includegraphics[width=.65\hsize]{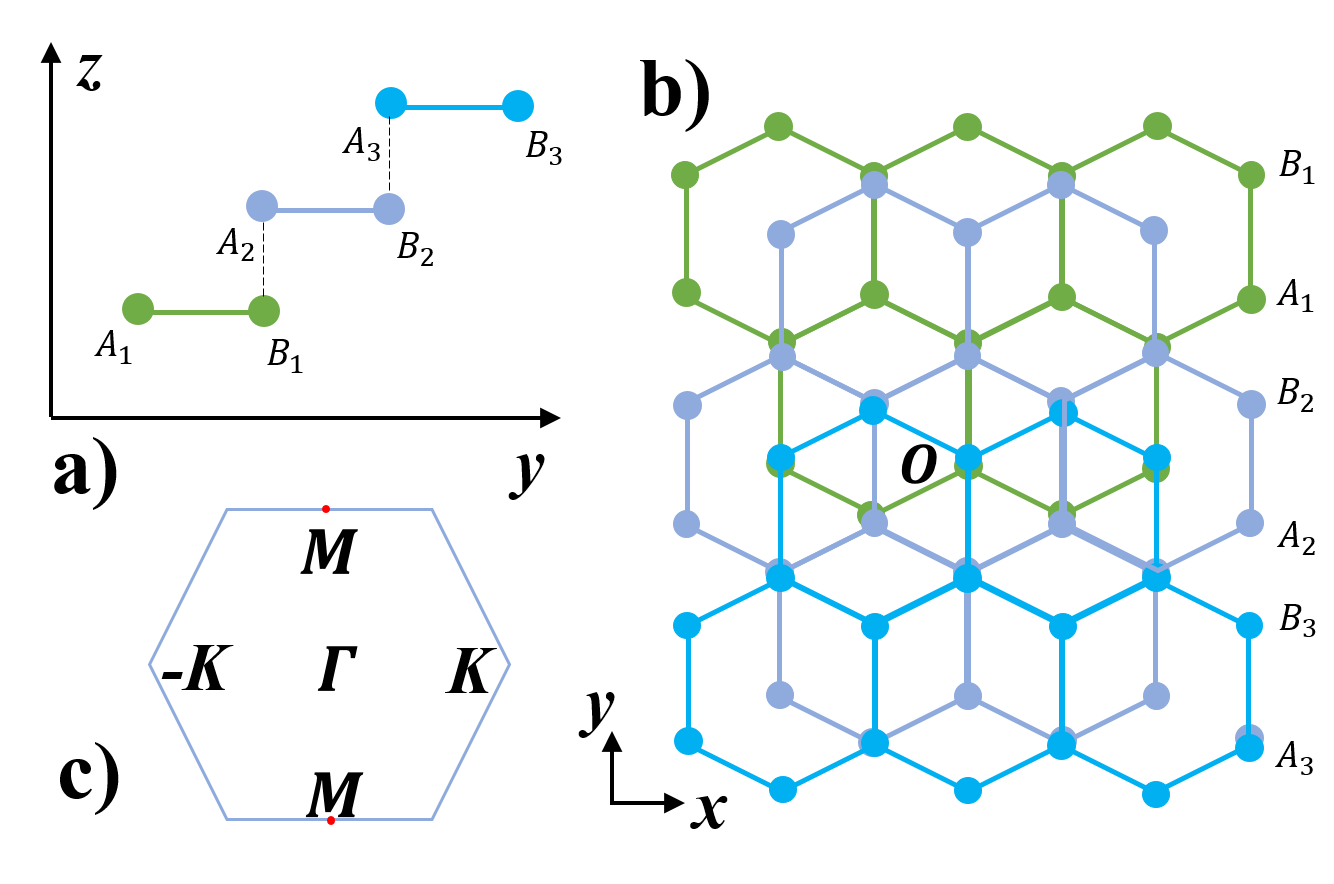}
\caption{
Trilayer Graphene with ABC stacking, denoted as ABC-TLG. a) Side view. b) Top view. The origin $O$ is chosen as the hexagon center of second layer, which is also the registered site of $A_1$ and $B_3$. c) Brillouin zone of b).
}
\label{fig1}
\end{figure}

Denote the lattice constants of ABC-TLG and hexagonal boron nitride (h-BN) as $ a_{G} $ and $ a_{BN} $ respectively. For simplicity, we assume there exist two coprime positive integers $ m,n $ such that $ ma_{G}=na_{BN}=L_{\rm M} $ is the superlattice constant. Denote $ \delta =(a_G -a_{BN})/a_{BN} $ as the lattice constant mismatch, then moir\'e superlattice constant is $ L_{\rm M}=a_{G}/|\delta| $.

The substrate h-BN will create a periodic potential on the moir\'e scale
\begin{eqnarray}
V_{\xi}(\bm r)=V_{\xi}\frac{1+\xi\sigma_{z}}{2}\sum_{j=0}^{2}\cos(\bm G_{j}\cdot\bm r+\phi_{\xi})+V\sigma_z
\end{eqnarray}
where $ \bm G_{j}=4\pi/(\sqrt{3}L_{\rm M})(-\sin\frac{2\pi j}{3},\cos\frac{2\pi j}{3}) $, and $\xi =\pm$ denotes different alignments of TLG and h-BN. 
\subsubsection{High-order van Hove singularity}
In this section we consider alignment scheme $\xi =-1$ where only B sublattice is affected by periodic potential of h-BN.
The moir\'e parameters are
\begin{eqnarray}
\delta =-0.017,\quad V_{-}=12.09{\rm meV},\quad\phi_{-}=-0.2591\pi.
\end{eqnarray}

We focus on the valence bands in Fig. \ref{fig}. Near $K'$ point, we can expand the dispersion of the upper valence band as
\begin{eqnarray}
E(\bm p-\bm K)=E_{\rm v}+\alpha p^2+\gamma (p_{x}^3-3p_{x}p_{y}^2)
\end{eqnarray}
where the coefficients $\alpha,\gamma$ and $\kappa$ are functions of electric field $V$. Near transition field $ V_c=0.06 $eV, the corresponding Fermi contours are shown in Fig. \ref{fig_tr}. We find that when $ V=V_c $, $K'$ point is a type-I high-order VHS where second order coefficients all vanish and three Fermi surfaces merge together.

\begin{figure}
\centering
\includegraphics[width=1\hsize]{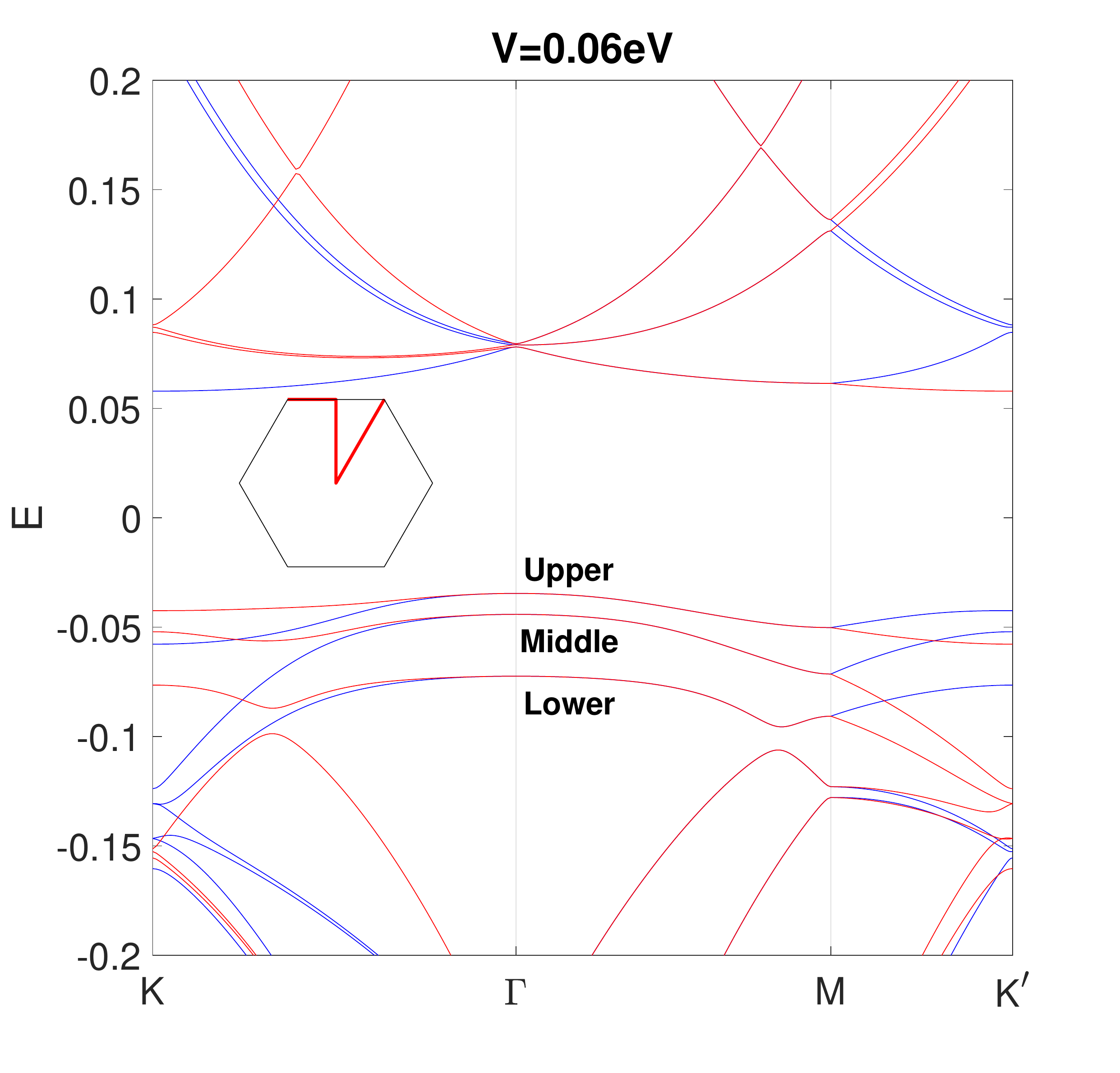}
\caption{
Band structure at a specific displacement field strength, where blue and red colors denote different valleys, and upper, middle and lower bands are specified.
}
\label{fig}
\end{figure}

\begin{figure*}
\centering
\includegraphics[width=1.05\hsize]{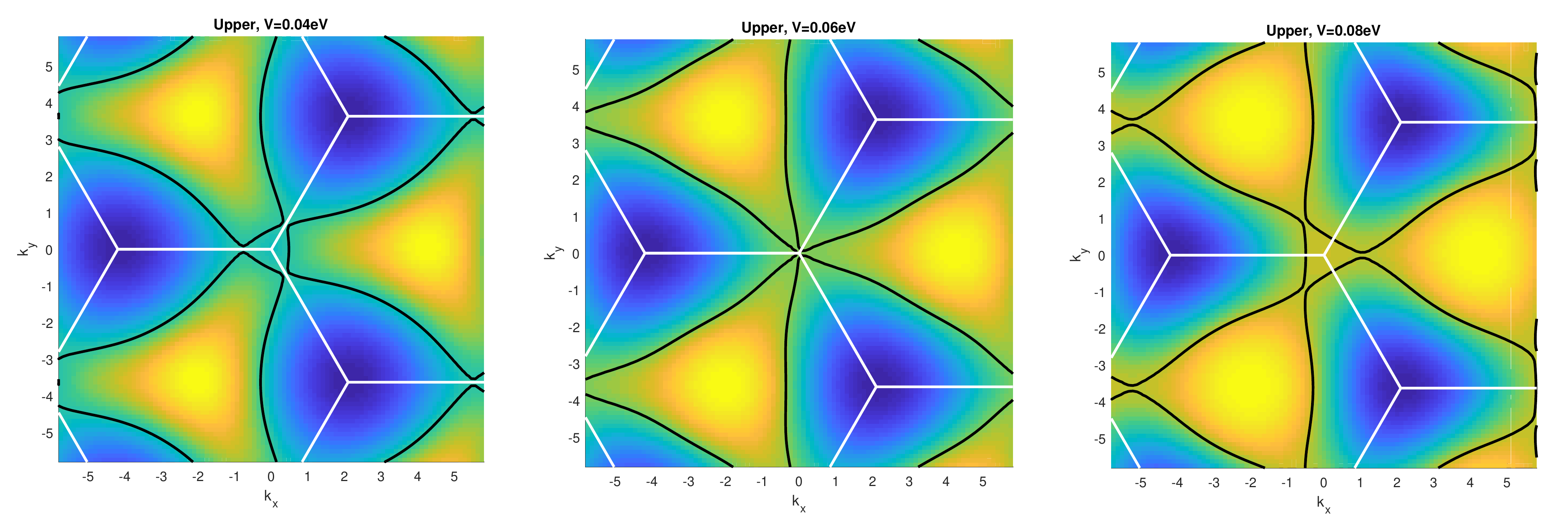}
\caption{
Topological transition as electric field changes.
}
\label{fig_tr}
\end{figure*}

\subsubsection{Effective tight-binding model}
We use the following tight-binding model to fit the two trivial bands in the continuum model spectrum
\begin{eqnarray}
H_{tb}=\sum_{\langle ij\rangle}t_{1}{c}_{i}^{\dagger}e^{i\phi\sigma_{z}}{c}_{j}+\sum_{\langle\langle ij\rangle\rangle}t_{2}{c}_{i}^{\dagger}{c}_{j}+h.c.
\end{eqnarray}

For moir\'e parameters in alignment scheme $\xi =+1$ where only A sublattice is affected by periodic potential of h-BN
\begin{eqnarray}
\delta =-0.017,\quad V_{+}=-14.88{\rm meV},\quad\phi_{+}=0.2788\pi,\quad V=70{\rm meV},
\end{eqnarray}
the corresponding fitting parameters of the lowest conduction bands are
\begin{eqnarray}
t_1=1.20{\rm meV},\quad\phi=-0.48\pi,\quad t_2=0.45{\rm meV}.
\end{eqnarray} 
Since $ \phi $ is close to $\pi/2$, the nearest neighbor term mostly contributes to the warping effect.
Fittings both along specific path and in the entire MBZ are shown in Figs. \ref{fig4} and \ref{fig3} respectively.

\begin{figure}
\centering
\includegraphics[width=1\hsize]{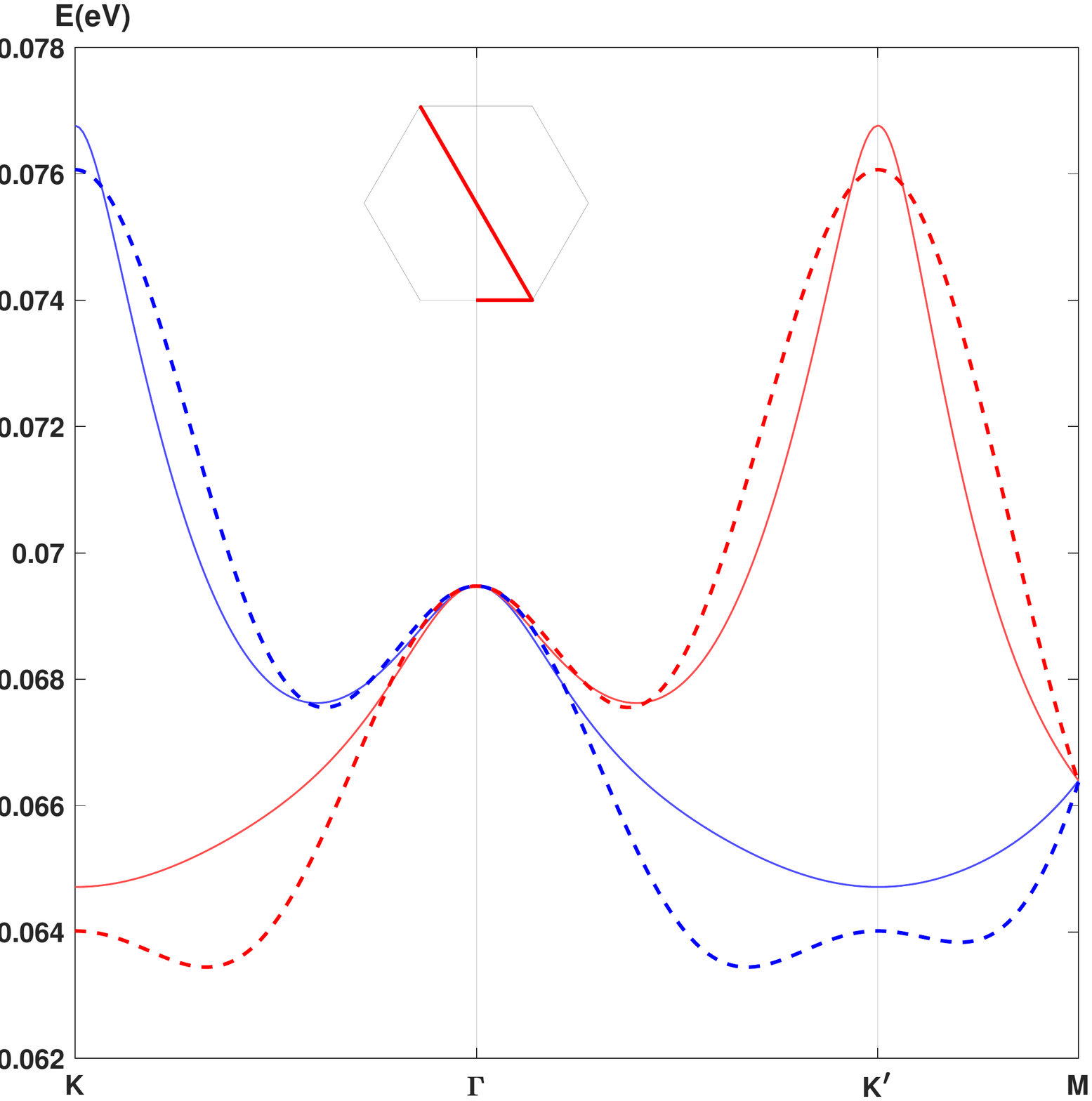}
\caption{
Tight-binding fitting (dashed) of the continuum model band structure (solid) along the path indicated by red lines in the hexagon. Red and blue denote different valleys.
}
\label{fig4}
\end{figure}

\begin{figure}
\centering
\includegraphics[width=1\hsize]{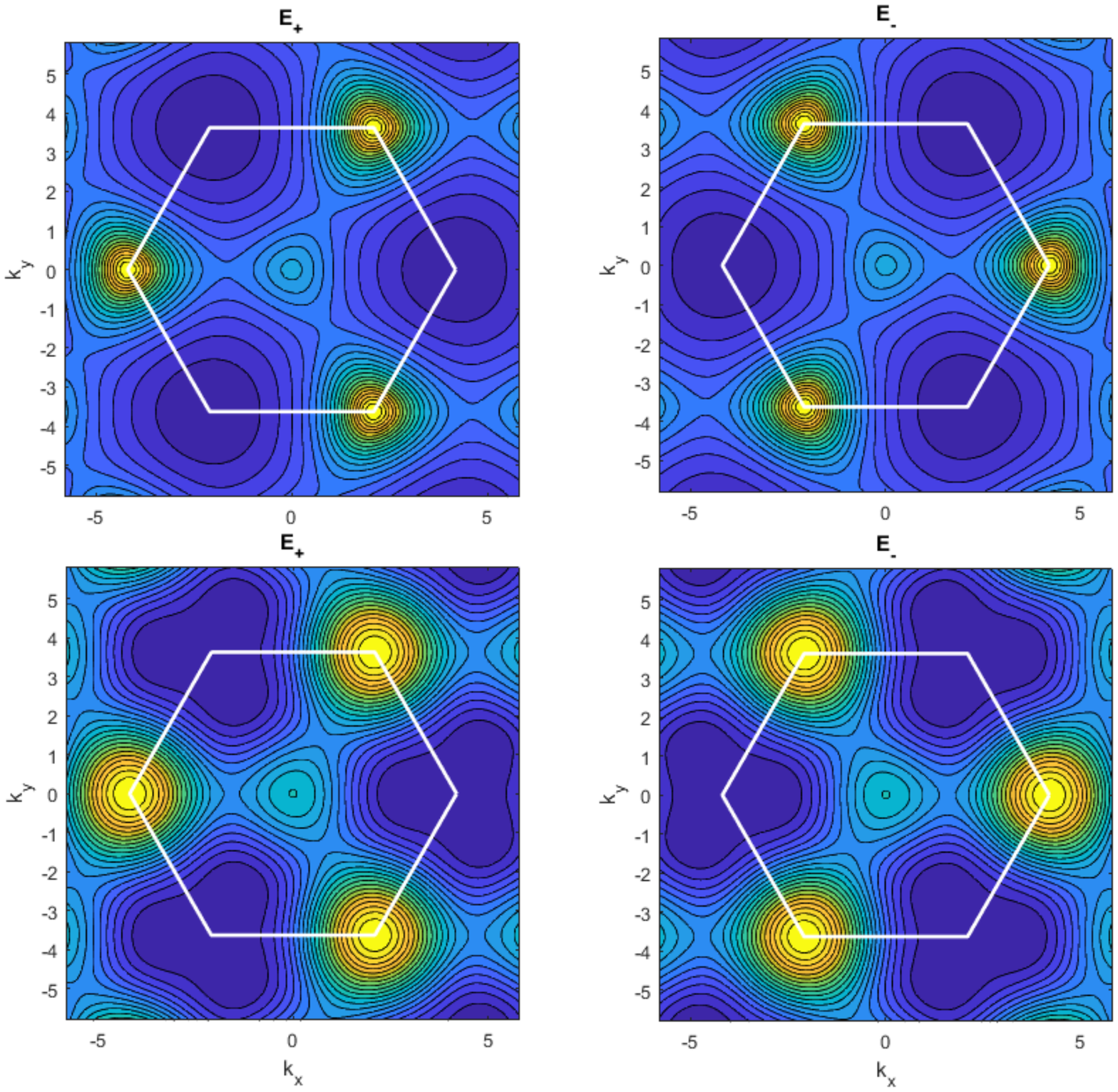}
\caption{
Energy contours calculated from continuum model (upper panels) and fitting of tight-binding model (lower panels) for different valleys denoted by $\pm$.
}
\label{fig3}
\end{figure}
\end{widetext}
\end{document}